\documentclass[journal]{IEEEtran}
\usepackage{cite}
\usepackage{amsmath,amssymb,amsfonts}
\usepackage{algorithmic}
\usepackage{graphicx}
\usepackage{textcomp}
\usepackage{times}
\usepackage{bm}
\usepackage{multirow}
\usepackage{enumerate}
\usepackage{graphicx}
\usepackage{subfigure} 
\usepackage{multirow}
\usepackage{algorithmic}
\usepackage{algorithm}
\usepackage{caption}
\usepackage[dvipsnames,usenames]{color}
\usepackage[usenames,dvipsnames]{color}
\usepackage{array}
\usepackage{subeqnarray}
\usepackage{cases}
\usepackage{stfloats}
\definecolor{tabcolor}{rgb}{1,0,0}
\DeclareCaptionLabelSeparator{twospace}{.\ ~}
\newcommand{\tabincell}[2]{\begin{tabular}{@{}#1@{}}#2\end{tabular}}  
\def\BibTeX{{\rm B\kern-.05em{\sc i\kern-.025em b}\kern-.08em
		T\kern-.1667em\lower.7ex\hbox{E}\kern-.125emX}}
\begin{document} 
\title{A Survey of Blockchain and Artificial Intelligence for 6G Wireless Communications}
\author{Yiping Zuo,~\IEEEmembership{Member,~IEEE}, Jiajia Guo,~\IEEEmembership{Member,~IEEE}, Ning Gao, ~\IEEEmembership{Member,~IEEE}, \\ Yongxu Zhu,~\IEEEmembership{Senior Member,~IEEE}, Shi Jin,~\IEEEmembership{Senior Member,~IEEE}, and Xiao Li,~\IEEEmembership{Member,~IEEE}
 
\thanks{Y. Zuo is with the College of Computer, Nanjing	University of Posts and Telecommunications, Nanjing 210028, China, and also with the National Mobile Communications Research Laboratory, Southeast University, Nanjing 210096, China (Email: zuoyiping@njupt.edu.cn).\\	
J. Guo, N. Gao, S. Jin, and X. Li are with the National Mobile Communications Research Laboratory, Southeast University, Nanjing 210096, China (Email: jiajiaguo@seu.edu.cn; ninggao@seu.edu.cn; jinshi@seu.edu.cn; li\_xiao@seu.edu.cn). \\
Y. Zhu is with the School of Engineering, University of Warwick, Coventry, UK (Email: yongxu.zhu@warwick.ac.uk).\\
(\textit{Corresponding author: Shi Jin)}} 
}

\maketitle
\begin{abstract}
The research on the sixth-generation (6G) wireless communications for the development of future mobile communication networks has been officially launched around the world. 6G networks face multifarious challenges, such as resource-constrained mobile devices, difficult wireless resource management, high complexity of heterogeneous network architectures, explosive computing and storage requirements, privacy and security threats. To address these challenges, deploying blockchain and artificial intelligence (AI) in 6G networks may realize new breakthroughs in advancing network performances in terms of security, privacy, efficiency, cost, and more. In this paper, we provide a detailed survey of existing works on the application of blockchain and AI to 6G wireless communications. More specifically, we start with a brief overview of blockchain and AI. Then, we mainly review the recent advances in the fusion of blockchain and AI, and highlight the inevitable trend of deploying both blockchain and AI in wireless communications. Furthermore, we extensively explore integrating blockchain and AI for wireless communication systems, involving secure services and Internet of Things (IoT) smart applications. Particularly, some of the most talked-about key services based on blockchain and AI are introduced, such as spectrum management, computation allocation, content caching, and security and privacy. Moreover, we also focus on some important IoT smart applications supported by blockchain and AI, covering smart healthcare, smart transportation, smart grid, and unmanned aerial vehicles (UAVs). {{Moreover, we thoroughly discuss operating frequencies, visions, and requirements from the 6G perspective.}} We also analyze the open issues and research challenges for the joint deployment of blockchain and AI in 6G wireless communications. Lastly, based on lots of existing meaningful works, this paper aims to provide a comprehensive survey of blockchain and AI in 6G networks. We hope this survey can shed new light on the research of this newly emerging area and serve as a roadmap for future studies.
\end{abstract}

\IEEEpeerreviewmaketitle
  
\section{Introduction} \label{sec:Introduction}
From 2020, the fifth-generation (5G) wireless networks achieve large-scale commercial deployment worldwide. Academia, industry, and governments are now engaged in research and development of the sixth-generation (6G) wireless communication technology to meet the demands of future networks in 2030 and beyond \cite{dang2020should}. Compared with 5G networks, 6G networks will have ultra-high network speed, ultra-low communication delay, and wider coverage depth. 6G networks will fully share ultra-high frequency wireless spectrum resources such as millimeter waves, terahertz, and light waves. 6G networks will also integrate technologies such as terrestrial mobile communications, satellite Internet, and microwave networks to form an integrated green network with group collaboration of all things, intelligent data perception, real-time security assessment, and coordinated coverage of space and earth \cite{yang20196g,tariq2020speculative,jiang2021road}. Facing the 6G era, the network will usher in new application scenarios and new performance requirements. In the 6G era, an air-space-ground integrated network communication system will be built to realize a ubiquitous network for full coverage and all scenarios \cite{zhang20196g}. However, diverse applications and communication scenarios, ultra-heterogeneous network connections, and service requirements for extreme performance all place higher requirements on the bandwidth, latency, security, connection density, and flexibility of 6G networks \cite{nguyen20216g,saad2019vision,chowdhury20206g}.

In the 6G era, artificial intelligence (AI) \cite{kato2020ten} is becoming more and more important. AI relies on mining big data for training and learning, continuously enhancing computing power to cope with higher transmission rates, and gaining more flexibility through continuous learning. In the future, 6G networks need to deal with explosive data traffic growth and massive device connections. Real-time management and control of these massive data will result in high complexity and latency overhead. Therefore, how to effectively perceive service characteristics, accurately monitor and control network resources, and dynamically allocate wireless resources has become an important issue for 6G networks. The use of AI at the application layer and network layer of 6G networks makes the network more intelligent and automated, which will be a necessary way to manage and control massive wireless big data \cite{yang2020artificial,khan2022efficient}. In addition, 6G network supports large-scale users, large-scale antennas, and multi-band hybrid transmission, so traditional physical layer transmission technologies will face multiple challenges in performance, complexity, and efficiency. This provides the possibility for AI technology to be applied to the wireless physical layer \cite{guo2022user,guo2021canet}. Notably, a 6G white paper \cite{latva2020key} released by the University of Oulu believed that AI will play an important role in 6G networks. The report of \cite{ITU-T2020framework} also indicated that it is necessary to introduce AI technology into complex network architectures in the future.

Blockchain is another highly anticipated emerging technology. In fact, blockchain is a technical system that integrates various technologies such as chain data structure, point-to-point transmission, distributed storage, consensus mechanism, and encryption algorithm \cite{leng2020blockchain}. The performance index requirements of 6G networks, such as ultra-high peak rate, ultra-low latency, ultra-high reliability, ultra-low energy consumption, and seamless connection, make system security, data privacy, sustainability, scalability, and other aspects subject to many risks and challenges \cite{wang2020security,conti2018survey}. Blockchain technology is an important technical means to cope with these challenges, especially with the advantages of distributed network architecture, intelligent node consensus, and smart contracts. The integrated application of blockchain and 6G networks provide a safe, intelligent, and efficient underlying technical support for the realization of the 6G network vision \cite{hewa2020role,zhu2019blockchain}. In particular, 6G white paper \cite{latva2020key} pointed out that 6G network requires an endogenous trust network, and blockchain technology may play an important role in the 6G networks to deal with a variety of complex new privacy challenges. Blockchain is a potential solution for privacy protection of 6G networks \cite{ylianttila20206g}. Moreover, blockchain can provide a strong guarantee for 6G networks to build a distributed, secure, and trusted transaction environment.

Research institutions and operators worldwide are accelerating the development of the cross-integration between emerging technologies such as blockchain and AI with 6G networks. The IMT-2030 (6G) Promotion Group's white paper proposes various scenarios for the application of blockchain technology in 6G networks, including dynamic spectrum management, ubiquitous access management, edge computing, and so on. China has established several international standard projects of blockchain in ITU, such as the establishment of ``Framework of blockchain of things as decentralized service platform" in ITU-T SG20 and the ``Reference framework for distributed ledger technologies" in ITU-T SG16. Sprint, an American operator, has partnered with NXM Labs to launch a 5G connected vehicle platform powered by blockchain technology. China Mobile and Huobi China have created a ``blockchain + Internet of Things (IoT)" identity authentication platform. Meanwhile, 3GPP specifically defines the network data analysis function, aiming to provide a standard interface for the development and application of AI models in wireless networks. The European Telecommunications Standards Institute has also established an industry standard working group to use AI for network management, expecting to achieve a high-level autonomous network with endogenous AI. The IMT-2030 (6G) Promotion Group puts forward the 6G vision of ``Intelligent Internet of Everything, Digital Twin", pointing out that 6G will enable the efficient and intelligent interconnection of all things.

\begin{table*}[!t] 
	\centering
	\caption {Comparison of our work with existing related research.} 
	\begin{tabular}{c|c|c|c|c|c|c|c}
		\hline \hline
		\tabincell{c}{\bf Research Work} &
		\tabincell{c}{\bf Year} &
		\tabincell{c}{\bf Blockchain \\ \bf for 5G/6G} &
		\tabincell{c}{\bf AI for \\ \bf 5G/6G} &
		\tabincell{c}{\bf Blockchain \\ \bf for AI} &
		\tabincell{c}{\bf AI for \\ \bf Blockchain} & 
		\tabincell{c}{\bf Blockchain and \\ \bf AI for 5G/6G} &  
		\tabincell{c}{\bf Key Technologies}
		\\ \hline 
		Wang et al. \cite{wang2021blockchain}                          
		&
		2021 
		& 
		Yes 
		&  
		No 
		&
		No 
		&
		No 
		&
		No 
		&
		Blockchain, RAN
		\\ \hline
		Wu et al. \cite{wu2019a}  
		& 
		2019 
		& 
		Yes 
		&  
		No
		&
		No 
		&
		No
		&
		Limited
		&
		Blockchain, IoT
		\\ \hline
		Nguyen et al. \cite{nguyen2020blockchain}
		& 
		2020
		& 
		Yes
		& 
		No 
		&
		No
		&
		No
		&
		Limited
		&
		Blockchain, IoT, SDN, NFV
		\\ \hline
		Yue et al. \cite{yue2021a}  
		& 
		2021
		& 
		Yes
		& 
		No
		&
		No
		&
		No
		&
		No
		&
		Blockchain, DApps
		\\ \hline
		Tahir et al. \cite{tahir2020review} 
		& 
		2020
		& 
		Yes
		& 
		No
		&
		No
		&
		No
		&
		No
		&
		Blockchain, RAN, D2D, SDN
		\\ \hline
		Bhat et al. \cite{bhat2020edge}  
		& 
		2020
		& 
		Yes
		& 
		No
		&
		No
		&
		No
		&
		Limited
		&
		Blockchain, IoT, MEC
		\\ \hline
		Sharma et al. \cite{sharma2021role} 
		& 
		2021
		& 
		Limited
		& 
		Yes
		&
		No
		&
		No
		&
		No
		&
		ML, DL, IoT, Blockchain
		\\ \hline
		Sun et al. \cite{sun2020machine} 
		& 
		2020
		& 
		Limited 
		& 
		Yes
		&
		No
		&
		No
		&
		No
		&
		ML, FL, Blockchain
		\\ \hline
		Rekkas et al. \cite{rekkas2021machine}  
		& 
		2021
		& 
		No
		& 
		Yes
		&
		No
		&
		No
		&
		No
		&
		ML
		\\ \hline
		Liu et al. \cite{liu2020federated} 
		& 
		2020
		& 
		No
		& 
		Yes
		&
		No
		&
		No
		&
		No
		&
		ML, FL
		\\ \hline
		Lin et al. \cite{lin2020artificial} 
		& 
		2020
		& 
		No
		& 
		Yes
		&
		No
		&
		No
		&
		Limited
		&
		AI, Blockchain
		\\ \hline
		{Salah et al. \cite{salah2019blockchain}} 
		& 
		{2019}
		& 
		{No}
		& 
		{No}
		&
		{Yes}
		&
		{No}
		&
		{No}
		&
		{Blockchain, AI}
		\\ \hline
		Shafay et al. \cite{shafay2022blockchain}  
		& 
		2022
		& 
		No
		& 
		No
		&
		Yes
		&
		No
		&
		No
		&
		Blockchain, DL, ML, FL
		\\ \hline
		Wang et al. \cite{wang2021applications}  
		& 
		2021
		& 
		No
		& 
		No
		&
		Yes
		&
		No
		&
		No
		&
		Blockchain, AI
		\\ \hline
		Xing et al. \cite{xing2018synergy}  
		& 
		2018
		& 
		No
		& 
		No
		&
		No
		&
		Yes
		&
		No
		&
		Blockchain, AI
		\\ \hline
		Pandl et al. \cite{pandl2020convergence}  
		& 
		2020
		& 
		No
		& 
		No
		&
		Yes
		&
		Yes
		&
		No
		&
		Blockchain, AI
		\\ \hline
		Dinh et al. \cite{dinh2018ai}  
		& 
		2018
		& 
		No
		& 
		No
		&
		Yes
		&
		Yes
		&
		No
		&
		Blockchain, AI
		\\ \hline
		Singh et al. \cite{singh2020blockiotintelligence} 
		& 
		2020
		& 
		No
		& 
		No
		&
		Yes
		&
		Yes
		&
		Limited
		&
		Blockchain, AI, IoT
		\\ \hline
		{Hussain et al. \cite{hussain2021artificial}}                           
		&
		{2021} 
		& 
		{No}
		&  
		{No} 
		&
		{No}
		&
		{Yes}
		&
		{No}
		&
		{Blockchain, AI}
		\\ \hline
		{Yang et al. \cite{yang2022fusing}}
		& 
		{2022}
		& 
		{No} 
		& 
		{No} 
		&
		{Limited} 
		&
		{Limited} 
		&
		{No} 
		&
		{Blockchain, AI, Metaverse}  
		\\ \hline
		{Mohanta et al. \cite{mohanta2020survey}} 
		& 
		{2020} 
		& 
		{No}
		&  
		{No}
		&
		{No} 
		&
		{No}
		&
		{Limited}
		&
		{Blockchain, AI, ML, IoT}
		\\ \hline
		{Tagde et al. \cite{tagde2021blockchain}}
		& 
		{2021}
		& 
		{Limited}
		& 
		{Limited} 
		&
		{No} 
		&
		{No} 
		&
		{Limited}
		&
		{Blockchain, AI}
		\\ \hline
		{Gill et al. \cite{gill2019transformative}}
		& 
		{2019} 
		& 
		{No}
		& 
		{No} 
		&
		{No} 
		&
		{No} 
		&
		{Limited}
		&
		{Blockchain, AI, IoT} 
		\\ \hline
		{Dhar et al. \cite{dhar2021blockchain}}
		& 
		{2021}
		& 
		{Limited} 
		& 
		{No} 
		&
		{No} 
		&
		{No}
		&
		{Limited}
		&
		{Blockchain, AI, IoT} 
		\\ \hline
		Liu et al. \cite{liu2020blockchain}  
		& 
		2020
		& 
		Yes
		& 
		Yes
		&
		Yes
		&
		Yes
		&
		Limited
		&
		Blockchain, ML
		\\ \hline
		Jameel et al. \cite{jameel2020reinforcement}  
		& 
		2020
		& 
		No
		& 
		No
		&
		No
		&
		Limited
		&
		Limited
		&
		Blockchain, RL, IIoT
		\\ \hline
		Wu et al. \cite{wu2021deep} 
		& 
		2021
		& 
		No
		& 
		No
		&
		No
		&
		Limited
		&
		Limited
		&
		Blockchain, DRL, IoT
		\\ \hline
		Miglani et al. \cite{miglani2021blockchain} 
		& 
		2021
		& 
		No
		& 
		No
		&
		Yes
		&
		Yes
		&
		Limited
		&
		Blockchain, ML, DL, RL, FL
		\\ \hline
		El Azzaoui et al. \cite{el2020block5gintell}  
		& 
		2020
		& 
		Yes
		& 
		Yes
		&
		No
		&
		No
		&
		Limited
		&
		Blockchain, AI, IoT
		\\ \hline
		Dibaei et al. \cite{dibaei2021investigating}
		& 
		2022
		& 
		Limited
		& 
		Limited
		&
		No
		&
		No
		&
		Limited
		&
		Blockchain, ML, DL
		\\ \hline
		\bf  Our Work
		& 
		2023
		& 
		\bf Yes
		&  
		\bf Yes 
		&
		\bf Yes
		&
		\bf Yes
		&
		\bf Yes
		&
		Blockchain, AI, IoT
		\\ \hline \hline
	\end{tabular}
	\label{tab:ComparisonOurWorkExistingResearch}
\end{table*}

\begin{table}[!t] 
	\centering
	\caption {List of major acronyms.} 
	\begin{tabular}{c|c}
		\hline \hline
		\tabincell{c}{\bf Acronyms} &
		\tabincell{c}{\bf Definitions} 
		\\ \hline
		AI
		& 
		Artificial Intelligence
		\\ \hline
		6G
		&
		Sixth-Generation
		\\ \hline
		IoT
		& 
		Internet of Things
		\\ \hline
		IIoT
		&
		Industrial Internet of Things
		\\ \hline
		UAV
		& 
		Unmanned Aerial Vehicle
		\\ \hline
		PoW
		&
		Proof of Work
		\\ \hline
		PoS
		&
		Proof of Stake
		\\ \hline
		DPoS
		&
		Delegated Proof of Stake
		\\ \hline
		PBFT
		&
		Practical Byzantine Fault Tolerance
		\\ \hline
		P2P
		&
		Peer-to-Peer
		\\ \hline
		MBS
		&
		Macro Base Station
		\\ \hline
		MEC
		&
		Mobile Edge Computing
		\\ \hline
		ML
		&
		Machine Learning
		\\ \hline
		DL
		&
		Deep Learning
		\\ \hline
		RL
		&
		Reinforcement Learning
		\\ \hline
		FL
		&
		Federated Learning
		\\ \hline
		DRL
		&
		Deep Reinforcement Learning
		\\ \hline
		KNN
		&
		K-Nearest Neighbor
		\\ \hline
		DNN
		&
		Deep Neural Network
		\\ \hline
		CNN
		&
		Convolutional Neural Network
		\\ \hline
		RNN
		&
		Recurrent Neural Network
		\\ \hline
		GAN
		&
		Generative Adversarial Network
		\\ \hline
		SGD
		&
		Stochastic Gradient Descent
		\\ \hline
		CSI
		&
		Channel State Information
		\\ \hline
		FDM
		&
		Frequency Division Multiplexing
		\\ \hline
		LSTM
		&
		Long Short-Term Memory 
		\\ \hline
		MIMO
		&
		Multiple-Input Multiple-Output
		\\ \hline
		RF
		&
		Radio Frequency
		\\ \hline
		MMSE
		&
		Minimum Mean Square Error
		\\ \hline
		AMP
		&
		Approximate Message Passing
		\\ \hline
		SDN
		&
		Software-Defined Networking
		\\ \hline
		RAN
		&
		Radio Access Network
		\\ \hline
		NFV
		&
		Network Function Virtualization
		\\ \hline
		DApps
		&
		Decentralized Applications
		\\ \hline
		D2D 
		&
		Device-to-Device
		\\ \hline
		BP
		&
		Belief Propagation
		\\ \hline
		RSU
		&
		Road Side Unit
		\\ \hline
		HDPC
		&
		High Density Parity Check
		\\ \hline
		LEO
		&
		Low Earth Orbit
		\\ \hline \hline
	\end{tabular}
	\label{tab:ListMajorAcronyms}
\end{table} 

\subsection{Previous Survey Works and Motivations} \label{subsec:PreviousWorksMotivations}
AI model or algorithm is based on the trained intelligence data. Meanwhile, blockchain is essentially a data storage method, or ``hyper ledger", which embodies data intelligence \cite{yuan2017blockchain}. Consequently, these two technologies, which are both closely related to data, can be effectively combined to complement each other and achieve technological improvement \cite{pandl2020convergence,dinh2018ai,singh2020blockiotintelligence}. As a trusted platform, blockchain can improve the authenticity, relevance, and validity of the data used by AI. From the perspective of data, computing power, and algorithms, blockchain improves the level of AI technology, innovates AI collaboration models and computing paradigms, and constructs a new AI ecosystem. With intelligent and automatic characteristics, AI can promote the natural evolution and data sorting of blockchain through the optimization and simulation of AI algorithms. Additionally, AI can effectively prevent the occurrence of blockchain node forks, can handle the operation of the blockchain more effectively, and improve efficiency intelligently. Most importantly, the close combination of blockchain and AI can promote and optimize various services and applications and also can provide a reliable, secure, and ultra-low latency intelligent network environment for next-generation wireless communications. Accordingly, in the future 6G networks, the research on the simultaneous deployment of blockchain and AI is of positive significance.

Next, we briefly describe the existing survey on the adoption of blockchain and AI in wireless communication systems. Researchers integrated blockchain with wireless communications to form secure and trusted mobile networks and services. In \cite{wang2021blockchain,wu2019a,nguyen2020blockchain,yue2021a,tahir2020review,bhat2020edge}, a large number of reviews on blockchain-supported wireless communications have been published, broadly elaborating the basic concepts, network architecture, enabling technologies, research challenges, and future research directions. Moreover, the mutual integration of blockchain and AI has also been investigated in detail by multiple studies \cite{salah2019blockchain,shafay2022blockchain,wang2021applications,xing2018synergy,pandl2020convergence,dinh2018ai,singh2020blockiotintelligence,hussain2021artificial,yang2022fusing}. Most importantly, the disruptive integration of blockchain and AI for wireless communications can greatly improve the network performance for a variety of services and applications. Many pieces of literature \cite{mohanta2020survey,tagde2021blockchain,gill2019transformative,dhar2021blockchain,liu2020blockchain,jameel2020reinforcement,wu2021deep,miglani2021blockchain,el2020block5gintell,dibaei2021investigating} have summarized and reviewed the topic of joint blockchain and AI for wireless communications. However, to the best of our knowledge, none of the existing surveys have comprehensively investigated this popular topic, especially few research emphasized the simultaneous deployment of blockchain and AI for next-generation wireless communications. For example, the research of \cite{liu2020blockchain} only briefly discussed the potential of the joint application of blockchain and machine learning (ML) in wireless communication systems. Similarly, the work in \cite{jameel2020reinforcement} briefly reviewed reinforcement learning (RL)-empowered blockchains applied in Industrial IoT (IIoT) networks. The authors of \cite{dibaei2021investigating} simply investigated the benefits of adopting blockchain with ML under the secure in-vehicle network. TABLE~\ref{tab:ComparisonOurWorkExistingResearch} displays a straightforward comparison of our work with existing related surveys.

\subsection{Novelty and Contributions} \label{subsec:NoveltyContributions}
Compared with the existing aforementioned works, our survey provides a comprehensive analysis and outlook on the current research progress of blockchain and AI for 6G wireless communications. We hope that this survey has some reference significance for carrying out more innovative research in this promising field. The main contributions of this article can be summarized as follows:
\begin{itemize}
	\item[1)] We briefly outline the basic knowledge of blockchain and AI. First, the concept, characteristics, and categories of blockchain and AI are introduced. Then, we separately discuss the classic applications of blockchain and AI for wireless communication systems.
	
	\item[2)] {{We systematically summarize the integration of blockchain and AI from two directions: blockchain-assisted AI and AI-assisted blockchain. Furthermore, we also emphasize the advantages of integrating blockchain and AI for wireless communication systems.}}
	
	\item[3)] We deeply elaborate on the latest developments of combining blockchain and AI in 6G secure services. We specifically focus on some of the most popular 6G secure services, such as spectrum management, computation allocation, content caching, and security and privacy. 
	
	\item[4)] We review the latest achievements of joint blockchain and AI in 6G IoT smart applications. We extensively discuss some important 6G IoT smart applications, including smart healthcare, smart transportation, smart grid, and unmanned aerial vehicle (UAV).
	
	\item[5)] {{On the basis of the comprehensive survey, we thoroughly discuss operating frequencies, visions, and requirements from the 6G perspective. We also propose some open issues and research challenges that need to be resolved for the applications of blockchain and AI to 6G wireless communications, and summarize several future research directions.}}
\end{itemize}

\subsection{Outline of the Survey} \label{subsec:StructurePaper}
\begin{figure*}[!t]
	\centering    
	\includegraphics[width=0.66\linewidth]{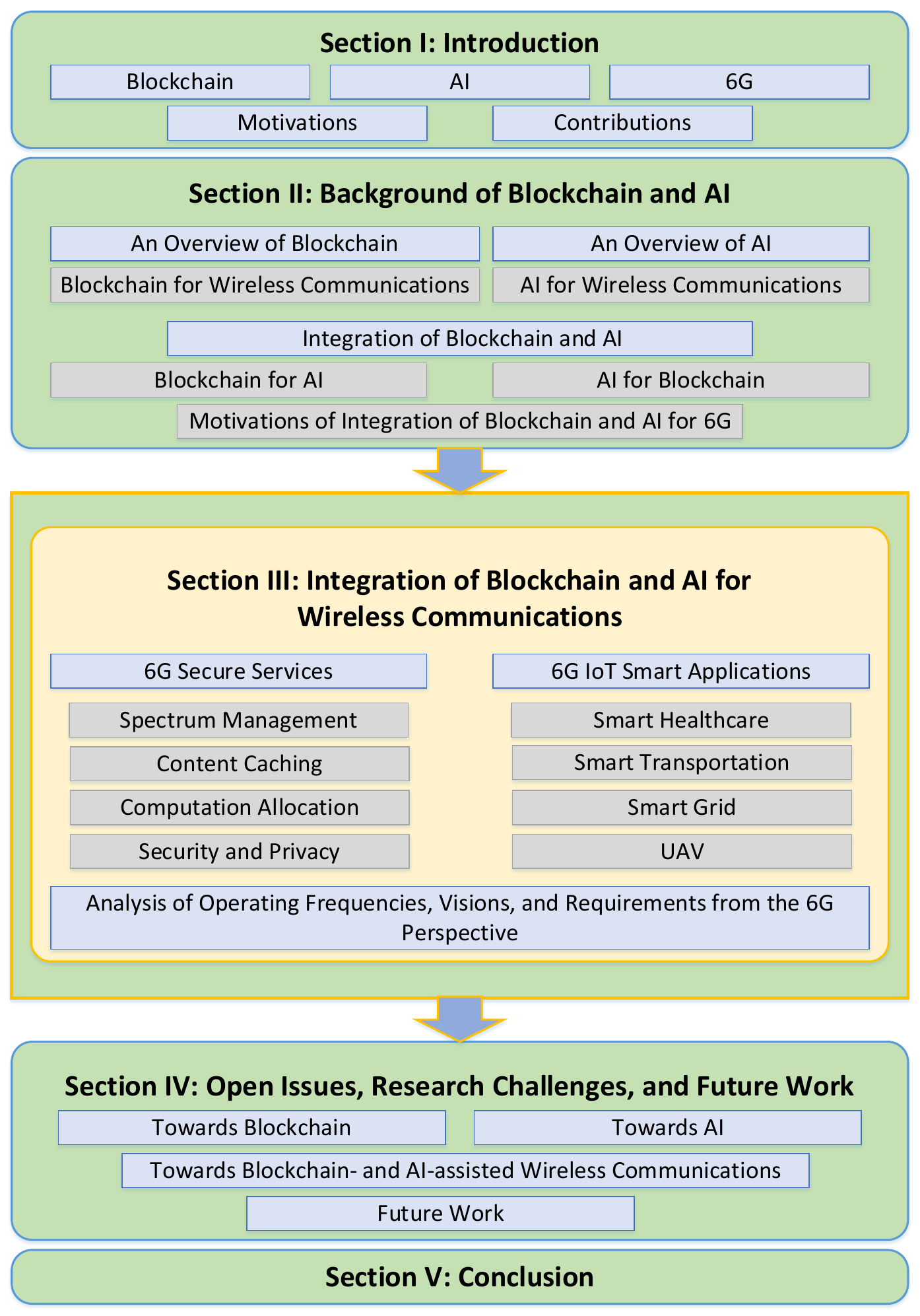}   
	\caption{{{Outline of the paper.}}}
	\label{fig:StructurePaper}
\end{figure*}

The outline of this article is presented in Fig.~\ref{fig:StructurePaper}. The remainder of this survey is organized as follows. Section \ref{sec:OverviewBlockchain} provides an overview of blockchain, including the concept, characteristics, categories, and representative applications in wireless communications. In Section \ref{sec:OverviewAI}, we describe an overview of AI, taking into account the concept, characteristics, categories, and typical applications in wireless communications. In Section \ref{sec:IntergrationBlockchainAI}, we discuss the mutual fusion of blockchain and AI, and emphasize the abundant benefits of this fusion for wireless communication systems. Section \ref{sec:IntegrationBlockchainAI6G} presents the integration of blockchain and AI for wireless communications, covering secure services and IoT smart applications. Some open issues and research challenges are discussed in Section \ref{sec:ResearchChallengs}, and the future work is also addressed. Finally, we conclude the main works of the survey in Section \ref{sec:ConclusionFutureWork}. The major acronyms used in this paper are summarized in TABLE~\ref{tab:ListMajorAcronyms}.

\section{{Background of Blockchain and AI}} \label{sec:BlockchainAI}
\subsection{{An Overview of Blockchain}} \label{sec:OverviewBlockchain}
\subsubsection{Concept of Blockchain} \label{subsec:ConceptBlockchain}
\begin{figure*}[!t]
	\centering    
	\includegraphics[width=1.0\linewidth]{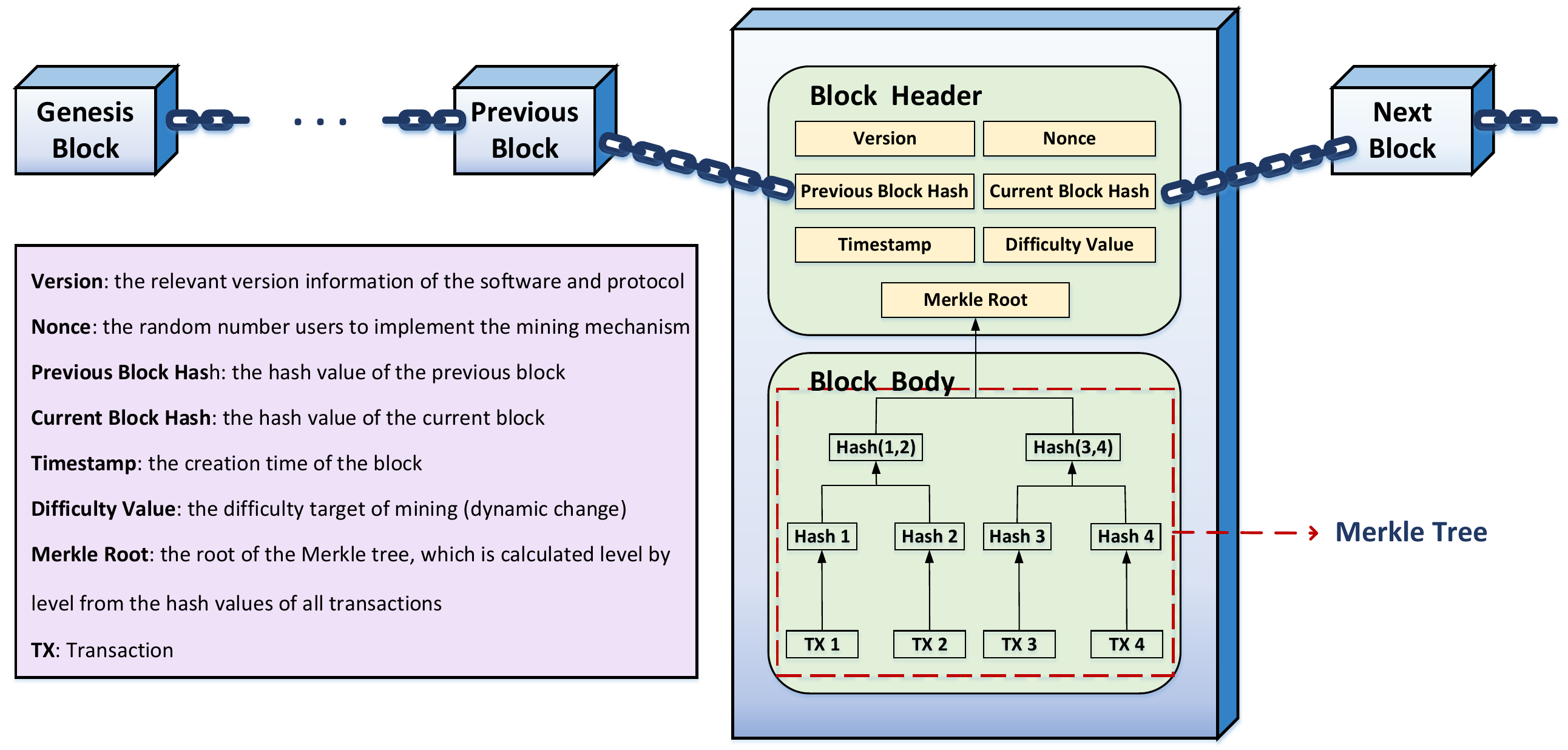}   
	\caption{The structure of blocks.}
	\label{fig:BlockStructure}
\end{figure*}

The concept of blockchain was first mentioned in the Bitcoin white paper written by S. Nakamoto \cite{nakamoto2008bitcoin}, marking the birth of blockchain 1.0. Ethereum based on smart contracts means the arrival of the blockchain 2.0 era \cite{aggarwal2021blockchain}. Blockchain 3.0 emphasizes its application to all aspects of society. Blockchain is essentially a distributed super-accounting ledger \cite{maesa2020blockchain}. This digital ledger guarantees data security through encryption algorithms and consensus mechanisms. Over time, the past transaction records on the blockchain ledger will not be deleted and cannot be tampered with. The blockchain network consists of multiple peer nodes, and these nodes do not need to trust each other. Each node independently maintains a copy of the global ledger. The transaction data in the ledger is encapsulated by blocks. The new block will be added to the end of the previous block in the form of a linked list, so this accounting ledger is called ``blockchain" \cite{Tschorsch2016bitcoin}. Taking bitcoin as an example, the typical block structure is shown in Fig.~\ref{fig:BlockStructure}. Each block is divided into two parts: block header and block body. The block body stores the verified transaction data. The block header contains the version, hash value of the previous block, hash value of the current block, timestamp, difficulty value, nonce, and Merkle root.

In the blockchain system, the consensus algorithm ensures that each node can maintain the same transaction content and sequence, which is the core component of the blockchain network \cite{wang2019asurvey}. Currently, the widely used and common consensus algorithms are as follows: Proof of Work (PoW) \cite{nakamoto2008bitcoin}, Proof of Stake (PoS) \cite{kiayias2017ouroboros}, Delegated Proof of Stake (DPoS) \cite{larimer2014delegated}, and Practical Byzantine Fault Tolerance (PBFT)\cite{castro2002practical}. PoW introduces distributed node computing power competition to maintain data consistency and consensus security. The core idea of PoS is that the node with the highest stake obtains the accounting right of the block. DPoS elects representatives through shareholder voting to get the right to keep accounts. PBFT sorts the request through the leader node, the follower node responds to the request, and the response result of most nodes is the final result. In addition, there is no perfect consensus protocol, because the adopted consensus protocol needs to be matched according to the type of blockchain system used. These algorithms have their own advantages but also have their own shortcomings, as presented in TABLE~\ref{tab:ComparisonConsensusAlgorithms}.

Since blockchain systems run in the {Peer-to-Peer (P2P)} network where nodes do not trust each other, the initiated transaction  needs to be completed under the witness of all nodes in the network. The transaction execution process of the blockchain is as follows. Specifically, the node first randomly generates its own private key and public key and constructs a transaction through a wallet or script tool, and then uses its own private key to sign the transaction. The signed transaction is propagated between neighbor nodes through the P2P network. Then, the node receiving the transaction verifies the legality of the transaction, and the miner digs out a new block according to the consensus algorithm. Next, the miners broadcast the new block to other nodes through the P2P network. Other miners verify the legitimacy of the new block to decide to discard or add to the local chain. After confirming the new block through the nodes of the whole network, this indicates that the new transaction is transferred successfully.

\subsubsection{Characteristics of Blockchain} \label{subsec:CharacteristicsBlockchain}
The development of blockchain technology has formed a relatively complete technology stack. Blockchain has been widely concerned and studied because of its important characteristics: decentralization, non-tampering, traceability, and anonymity \cite{wu2019a}.

{\bf Decentralization:} Blockchain technology is to complete data interaction without relying on any third-party intermediaries or institutions.  Compared with the centralized network, the bottom layer of blockchains adopts the P2P network architecture. In the blockchain network, there is no traditional central server to process data recording, storage, and updating. Every node is equal, and the data maintenance of the entire blockchain network is jointly participated by all nodes. In addition, the withdrawal of any node will not affect the operation of the entire system, and the blockchain network has strong robustness.

{\bf Non-Tampering:} Once the transaction data is packaged on the chain by miner nodes and permanently stored in the blockchain to form an immutable historical ledger. By storing the hash value of the previous block in each block, the blocks are connected back and forth to form a chain structure. This special chained data structure enables all blocks storing transaction data to be added to the end of the blockchain in chronological order. The malicious node wants to tamper with the data, which inevitably causes the hash value of the current block and all subsequent blocks to change, leading to the collapse of the chain structure. Therefore, the cost of data tampering becomes extremely high, making it almost impossible to modify the blockchain.

{\bf Traceability:} In blockchain networks, all transactions are public and any node can get a record of all transactions. Except for the encrypted private information of both parties to the transaction, all data on the blockchain can be queried through public interfaces. Blockchain uses the chain block structure with a timestamp to store data, resulting in adding a time dimension to data. Each transaction on the block is connected to two adjacent blocks through cryptographic methods, which guarantees that users can trace the source of any transaction.

{\bf Anonymity:} Since the nodes in the blockchain network do not need to trust each other, there is no need to disclose the identity between the nodes. This ensures the anonymity of each participating node in the blockchain system and protects the privacy of the nodes. Nodes can conduct transactions without knowing the identity of the other party. Both nodes of the transaction only need to publish their own addresses to communicate with each other. In the blockchain network, nodes use asymmetric encryption technology to build trust between nodes in an anonymous environment.

\subsubsection{Categories of Blockchain} \label{subsec:SmartContract}
According to different application scenarios, blockchains are classified into public blockchain, consortium blockchain, and private blockchain \cite{liu2020blockchain}.

{\bf Public Blockchain:} The public blockchain is a completely decentralized blockchain \cite{wood2014ethereum}, and any user can join the blockchain network. There is no official organization, management agency, and no central server. Participating nodes can freely enter and exit the network without being controlled.

{\bf Private Blockchain:} The private blockchain is fully centralized \cite{wu2019a}, and only authorized and trusted nodes can join the blockchain network. All nodes in the network are controlled by an organization. The operating rules and consensus mechanism of the system are determined by the organization itself.

{\bf Consortium Blockchain:} The consortium blockchain is a partially decentralized blockchain \cite{dib2018consortium} that is jointly maintained by multiple companies or organizations. This type of blockchain is between the public blockchain and the private blockchain and has the characteristics of multi-center or partial decentralization. Only members belonging to the alliance can generate transactions or view blockchain information.

According to the way of trust construction in different scenarios, the blockchain can also be divided into a permissionless blockchain and a permissioned blockchain \cite{nguyen2020blockchain}. The permissionless blockchain is also called the public blockchain, which is a completely open blockchain. That is, anyone can join the network and participate in the complete consensus accounting process. The permissioned blockchain is a semi-open blockchain. Only designated members can join the network, and each member has different rights to participate. The permission chain often establishes a trust relationship by issuing identity certificates. This blockchain has partial decentralization characteristics, which is more efficient than permissionless blockchains. Private blockchains and consortium blockchains belong to permission chains. The comparison of the characteristics of the above-mentioned different blockchains is shown in TABLE~\ref{tab:ComparisonCharacteristicsBlockchains}.

\subsubsection{Blockchain for Wireless Communications} \label{subsec:Blockchain6GNetworks}
Blockchain technology naturally has many advantages such as decentralization, traceability, distribution, and tamper resistance. Therefore, researchers expect to apply the blockchain to all levels of the wireless communication system, which will achieve a substantial increase in system performance and a true sense of the connection of everything \cite{nguyen2020blockchain}. Blockchain can provide traceable communication for 6G networks, which not only facilitates network administrators to query historical resource conditions at any time, but also reduces malicious users' behavior of fabricating resource usage. In addition, the blockchain uses multi-party consensus mechanisms to record the interactions between users, so as to ensure the fairness and openness of all interactions. The integration of blockchain and 6G will provide a strong security guarantee for the construction of a safe and credible communication ecosystem. So far, the research on the integration of blockchain and 6G network mainly involves two main aspects \cite{yue2021a}: blockchain-enabled 6G secure services \cite{zuo2021Blockchain1,ling2021practical,zuo2021computation,zuo2021delay,ling2021databroker,zuo2021blockchain2,zuo2020computationiccc,zuo2021computationvtc,kotobi2017blockchain,kotabi2018secure,bayhan2018spass,bayhan2019smart,weiss2019on,qiu2020blockchain,grissa2019trustas,maksymyuk2019blockchain,zhang2020ablockchain,jiang2021decentralized,li2019credit,chatzopoulos2018flopcoin,sun2020joint,wang2021consortium,kang2019blockchain,sun2020blockchain,xu2020blockchain,Fan2018BlockchainbasedEP,ei2019asingle,sun2019blockchainenabled,liu2020efficient} and blockchain-assisted 6G Internet of Things (IoT) smart applications \cite{singh2021anovel,nguyen2021bedgehealth,yazdinejad2020decentralized,xia2020abayesian,yin2021ablockchain,chattaraj2021blockclap,wang2020blockchain,bera2021designing,jindal2020guardian,gai2021blockchain,tan2021blockchainbased,jiang2020incentivizing}.

\begin{table*}[!t] 
	\centering
	\caption {Comparison of four typical consensus algorithms.} 
	\begin{tabular}{c|c|c|c|c|c|c|c|c}
		\hline \hline
		\tabincell{c}{\bf Consensus \\ \bf Algorithm} &
		\tabincell{c}{\bf Security} &
		\tabincell{c}{\bf Decentralization} &
		\tabincell{c}{\bf Fork} &
		\tabincell{c}{\bf Resource \\ \bf Consumption} & 
		\tabincell{c}{\bf Transaction Confir \\ \bf mation Time} &
		\tabincell{c}{\bf Transaction \\ \bf Throughput} &
		\tabincell{c}{\bf Network \\ \bf Scale} & 
		\tabincell{c}{\bf Typical Application \\ \bf System}           
		\\ \hline 
		\bf PoW                          
		&
		High
		& 
		Higher
		&  
		Easy
		&
		Large
		&
		Long
		&
		Small
		&
		Large
		&
		Bitcoin
		\\ \hline
		\bf PoS
		& 
		Higher
		& 
		Higher
		&  
		Easy
		&
		General
		&
		General
		&
		Small
		&
		Large
		&
		Peercoin
		\\ \hline
		\bf DPoS
		& 
		General
		& 
		Low
		& 
		Hard 
		&
		Small
		&
		Short
		&
		General
		&
		Large
		&
		Bitshares
		\\ \hline
		\bf PBFT
		& 
		General
		& 
		General
		&  
		Hard
		&
		Small
		&
		Short
		&
		General
		&
		Small
		&
		Hyperleder
		\\ \hline \hline
	\end{tabular}
	\label{tab:ComparisonConsensusAlgorithms}
\end{table*}

\begin{table*}[!t] 
	\centering
	\caption {Comparison of different blockchains.} 
	\begin{tabular}{c|c|c|c}
		\hline \hline
		\tabincell{c}{\bf Type of \\ \bf Blockchain} &
		\tabincell{c}{\bf Public Blockchain \\ \bf (Permissionless Blockchain)} &
		\tabincell{c}{\bf Private Blockchain \\ \bf (Permissioned Blockchain)} &
		\tabincell{c}{\bf Consortium Blockchain \\ \bf (Permissioned Blockchain)} 
		\\ \hline
		\bf Degree of Centralization
		& 
		Decentralization
		& 
		Centralization
		&  
		Multi-centralization
		\\ \hline
		\bf Participant
		& 
		Anyone 
		& 
		Designated member 
		& 
		Alliance member 
		\\ \hline
		\bf Bookkeeper 
		& 
		All participants 
		& 
		Self-determined 
		&  
		\tabincell{c}{Participants decided after negotiation} 
		\\ \hline
		\bf Advantage
		&
		High credibility
		&
		High security \& Low latency
		&
		Good scalability
		\\ \hline
		\bf Disadvantage
		&
		High latency \& Low efficiency
		&
		Limited nodes \& Centralization
		&
		Have a trust issue
		\\ \hline
		\tabincell{c}{\bf Typical Application \\ \bf Scenarios}
		&
		Bitcoin, Ethereum
		&
		Hyperledger
		&
		Centralized Exchange
		\\ \hline \hline
	\end{tabular}
	\label{tab:ComparisonCharacteristicsBlockchains}
\end{table*}

{\bf Secure Services:} The research on the blockchain-enabled 6G secure services mainly involves spectrum sharing \cite{kotobi2017blockchain,kotabi2018secure,bayhan2018spass,bayhan2019smart,weiss2019on,qiu2020blockchain,grissa2019trustas,maksymyuk2019blockchain,zhang2020ablockchain,jiang2021decentralized}, computing and storage 
\cite{li2019credit,chatzopoulos2018flopcoin,sun2020joint,wang2021consortium,kang2019blockchain,sun2020blockchain,xu2020blockchain,Fan2018BlockchainbasedEP}, interference management \cite{ei2019asingle,sun2019blockchainenabled,liu2020efficient}, and so on. Take blockchain based spectrum sharing as an example, the authors of \cite{kotobi2017blockchain} and \cite{kotabi2018secure} proposed a blockchain-based verification and access control protocol to complete the spectrum sharing between primary and secondary users. The works in \cite{bayhan2018spass,bayhan2019smart} proposed a blockchain-based spectrum sensing as a service solution. Here, the smart contract is mainly responsible for the following functions: 1) Scheduling the spectrum allocation between users and the helper to maximize system revenue; 2) Identifying whether the helper is a malicious node, and ensuring the security of spectrum sharing. \cite{weiss2019on} discussed the application of blockchain in different spectrum access scenarios and analyzed the advantages and disadvantages of different spectrum sharing mechanisms. Based on the consortium blockchain, a secure spectrum trading and sharing scheme for drone-assisted communication systems was contrived in \cite{qiu2020blockchain}. To solve the issue of privacy risk in spectrum sharing, \cite{grissa2019trustas} proposed a trusted framework found on blockchain entitled TrustSAS for dynamic spectrum access. The work in \cite{maksymyuk2019blockchain} introduced a smart network architecture, which uses smart contracts to handle unlicensed spectrum sharing between operators and users. The authors of \cite{zhang2020ablockchain} proposed a blockchain-based radio service model, which can reduce the management cost of dynamic access systems. For wireless downlink communication systems with multiple mobile operators, the work in \cite{jiang2021decentralized} delineated a blockchain-based dynamic spectrum acquisition scheme.

The computing and storage capabilities of edge computing are valuable network resources, which can be efficiently managed through the blockchain. To solve the problem of low efficiency of computing resource transactions, in the blockchain-based edge-assisted IoT network, the work in \cite{li2019credit} considered using the credit-based payment for fast computing resource transactions. The work in \cite{chatzopoulos2018flopcoin} proposed a blockchain-based multi-layer computing offloading architecture, which enhances the collaboration between users in sharing computing resources. In the blockchain-empowered multi-task cross-server edge computing scenario, the authors of \cite{sun2020joint} proposed two double auction mechanisms to drive end users and edge servers to securely allocate and trade resources. The works of \cite{wang2021consortium} and \cite{kang2019blockchain} applied the consortium blockchain and smart contracts to the vehicle edge computing network for resource trading, data storage, and data sharing, and to defend against malicious behaviors of vehicles. Blockchain was used to construct an attribute-based encryption scheme for secure storage and sharing of electronic medical records in \cite{sun2020blockchain}. The authors of \cite{xu2020blockchain} designed a blockchain-enabled arbitrable remote data auditing scheme to provide reliable network storage services. To deal with the privacy issues in content-centric mobile networks, the study in \cite{Fan2018BlockchainbasedEP} proposed a secure and efficient blockchain-inspired encrypted cloud storage solution.

The dense deployment of 6G networks will cause serious interference problems, so the use of blockchain for interference management is also a very important topic. The work in \cite{ei2019asingle} described a greedy distributed algorithm, using the blockchain currency mechanism and coordination protocol. This algorithm can realize the optimal information distribution achieved by the traditional central controller before, and eliminate the interference between users. The authors of \cite{sun2019blockchainenabled} analyzed the interference problem between transaction nodes in the blockchain-based IoT network, and derived the probability density function of the signal to interference plus noise ratio between IoT nodes and full nodes. The blockchain-based full node deployment solution can ensure a high transaction success rate and overall communication throughput, and protect the IoT network from security threats. In the blockchain-based femtocell network, to avoid excessive interference from femtocell users to the macro base station (MBS), the MBS set a price for the interference generated by the femtocell user in \cite{liu2020efficient}. Femtocell users determined their transmission power and payment fees according to the modeled Stackelberg game. Blockchain enabled the femtocell network to reliably make payments without the involvement of intermediaries.

{\bf IoT Smart Applications:} Blockchain has also been introduced into many IoT smart application systems, such as smart healthcare \cite{singh2021anovel,nguyen2021bedgehealth,yazdinejad2020decentralized}, smart transportation \cite{xia2020abayesian,yin2021ablockchain,chattaraj2021blockclap}, smart grid \cite{wang2020blockchain,bera2021designing,jindal2020guardian}, UAV \cite{gai2021blockchain,tan2021blockchainbased,jiang2020incentivizing}, and so on. For example, the authors of \cite{singh2021anovel} described a blockchain-energized patient-centric electronic medical record management architecture, and completed the prototype implementation of this architecture on the Hyperledger platform. The work in \cite{nguyen2021bedgehealth} proposed a mobile edge computing (MEC)- and blockchain-based distributed healthcare architecture for medical data offloading and data sharing. In the hospital network, the traditional centralized patient identity authentication method may cause problems such as long time and high cost. To resolve these problems, the authors of \cite{yazdinejad2020decentralized} designed a distributed patient authentication method using blockchain. A blockchain-enabled electricity trading scheme between vehicles was proposed in \cite{xia2020abayesian}. To alleviate the problem of incomplete information sharing, the Bayesian game was also used to price electricity. The study of \cite{yin2021ablockchain} delineated a blockchain storage system, which supports incremental data updates of vehicles. This system used multiple technologies such as data partitioning, smart contracts, and redundant backups. To deal with the security threats of the Internet of Vehicles, \cite{chattaraj2021blockclap} considered a blockchain-assisted certificateless key agreement protocol, which has high security and low communication and computing costs.

The integration of blockchain into smart grid and UAV is also a hot topic. In the smart grid system based on edge computing, to realize the private and secure communication between grid terminals and edge servers, the work of \cite{wang2020blockchain} introduced a blockchain-enabled anonymous authentication and key agreement protocol. In the IoT-supported smart grid system, the data of smart meters can be safely transmitted to service providers through the private blockchain-based access control protocol proposed in \cite{bera2021designing}. The work in \cite{jindal2020guardian} introduced a blockchain-empowered security demand response management scheme, which processes energy transaction requests generated in the smart grid system. The authors of \cite{gai2021blockchain} proposed a blockchain-energized multi-party authentication scheme, which can provide secure point-to-point wireless communication and reliable group communication for UAV networks. Under the heterogeneous UAV flight ad hoc network, the study of \cite{tan2021blockchainbased} drew up a blockchain-based distributed key management scheme, which includes four modules: cluster key distribution, key updating, cluster UAV migration, and malicious UAV revocation. In order to solve the security and privacy issues faced by energy micro-transactions, \cite{jiang2020incentivizing} introduced a distributed and secure UAV-assisted radio transmission architecture, which utilizes the directed acyclic graph and consortium blockchain.

\begin{figure*}[!t]
	\centering    
	\includegraphics[width=1.0\linewidth]{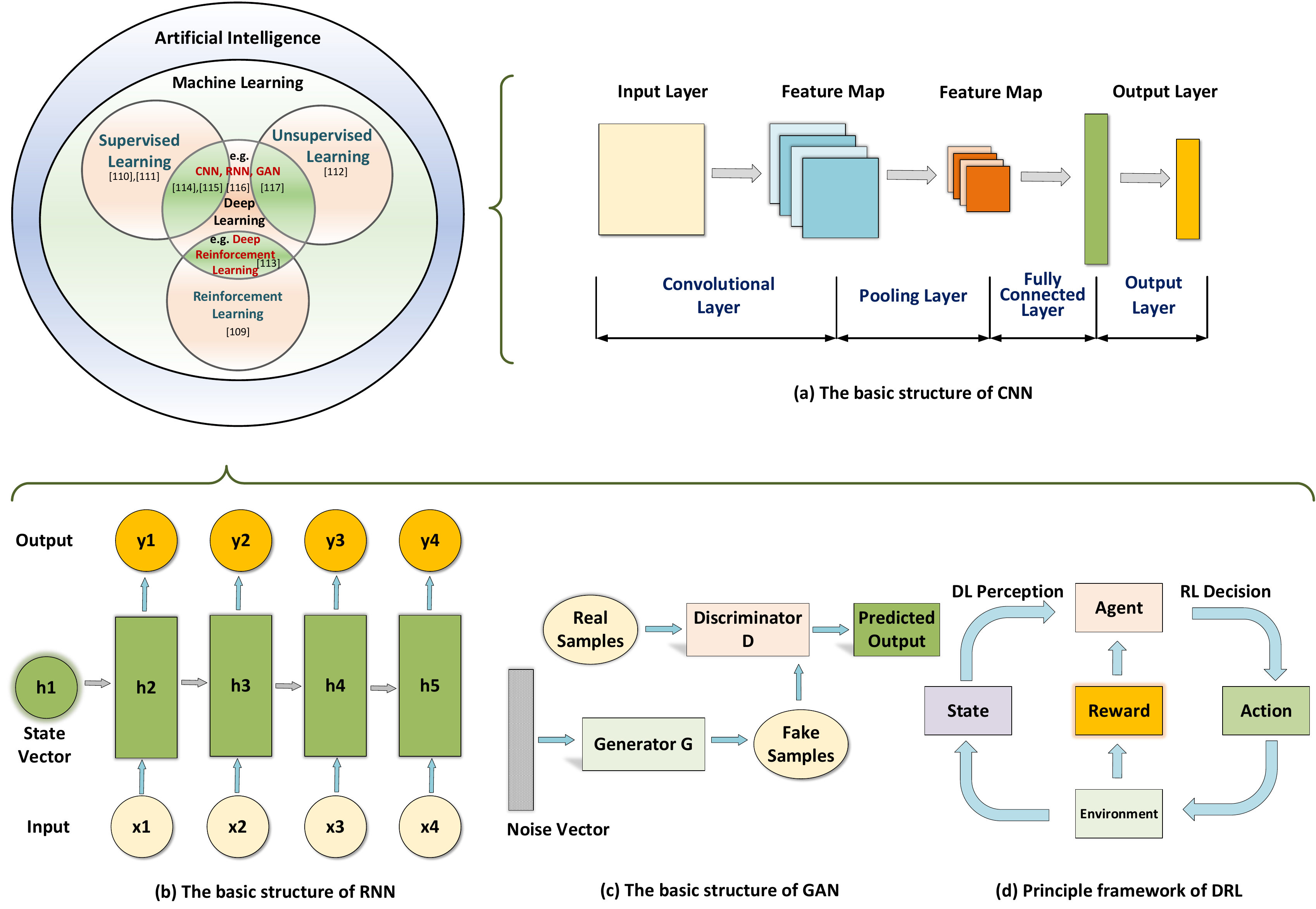}   
	\caption{{Classification of AI.}}
	\label{fig:ClassificationArtificialIntelligence}
\end{figure*}

\subsection{{An Overview of AI}} \label{sec:OverviewAI}
\subsubsection{Concept of AI} \label{subsec:ConceptAI}
Nowadays, AI has become a field with numerous practical applications and active research topics, and is booming\cite{guo2022overview}. It is difficult to give AI a scientific definition as rigorous as a mathematical one. Until now, ``what is AI ?" is still a debated issue in academia, and there is no unanimously accepted statement. Professor N. J. Nisson of Stanford University's AI Research Center believes that ``AI is the science of knowledge, that is, how to express knowledge, how to acquire knowledge, and how to use knowledge" \cite{nilsson2009quest}. Professor P. H. Winston of the Massachusetts Institute of Technology holds that ``AI is the study of how to make computers do intelligent jobs that only humans could do in the past" \cite{winston1992artificial}. From the perspective of knowledge engineering, Professor E. A. Feigenbaum of Stanford University considers that ``AI is a knowledge information processing system" \cite{barr1981handbook}. In a word, AI is a comprehensive discipline, which integrates many disciplines such as computer science, logic, biology, psychology, philosophy, etc. AI has achieved remarkable results in applications such as speech recognition, image processing, natural language processing, automatic theorem proving, and intelligent robots \cite{mao2018deep,lecun2015deep}.

In the early days of AI, problems that were very difficult for human intelligence but relatively simple for computers were dealt with quickly. For example, some problems can be described by a series of formal mathematical rules. The real challenge for AI lies in solving tasks that are easy for humans to perform but difficult to describe formally, such as recognizing words spoken by people or faces in images. For these problems, human beings can often solve them easily and intuitively. In recent years, most of the major breakthroughs in the field of AI can be summarized as the development and application of ML technology. The relationship between AI and ML is depicted in Fig.~\ref{fig:ClassificationArtificialIntelligence}. ML provides a solution for these relatively intuitive problems. This solution has its own ability to acquire knowledge, that is, the ability to extract patterns from raw data. Further, there is a key approach in ML, which can improve computer systems from experience and data. This approach allows computers to learn from experience and understand the world in terms of a hierarchical concept system, with each concept defined by its relationship to some relatively simple concepts. By allowing computers to acquire knowledge from experience,  human beings can avoid formally specifying all the knowledge they need. Hierarchical concepts let computers construct simpler concepts to learn complex concepts. Drawing a diagram representing how these concepts build on top of each other, we obtain a `deep' (many layers) diagram.
For this reason, we call this approach as deep learning (DL) \cite{lecun2015deep,zhou2022multiscale}. ML can build AI systems running in complex real-world environments. DL is a specific type of ML with great power and flexibility. In DL, the big world can be described as a nested hierarchical concept system. This hierarchical concept system refers to the definition of complex concepts by the connection between simpler concepts, and the generalization from general abstraction to high-level abstraction.

\subsubsection{Characteristics of AI} \label{subsec:CharacteristicsArtificial Intelligence}
In this subsection, we will discuss some important characteristics of AI, including data driving, uncertainty, environmental perception, and scalability.

{\bf Data Driving:} AI gradually completes the technology from artificial knowledge expression to big data-driven knowledge learning. AI rarely needs to rely on manual engineering, so it can easily take advantage of the increment in the amount of available computation and data \cite{goodfellow2016deep}. For example, a data-driven ML network regards the function to be implemented as an unknown black box, replaces it with an ML network, and then relies on a large amount of training data to complete the training from input to output.

{\bf Uncertainty:} There is a lot of uncertainty since AI has some similarities or differences compared with any other discipline such as mathematics, physics, cognitive, and behavioral psychology. Most areas of AI do not develop like traditional methods of mathematics, nor do they align with general models of physics. There will always be connections between AI and cognitive or behavioral psychology, but those connections ignore the mathematical and engineering themes.  As a prescience, the framework of AI is not yet complete.

{\bf Environmental Perception:} The AI system should be able to generate the ability to perceive the external environment with the help of sensors and other devices. AI can receive various information from the environment through hearing, vision, smell, and touch like humans, and generate necessary reactions to external input such as text, voice, expressions, and actions. These reactions even influence environmental or human decision-making. Ideally, an AI system should have certain adaptive characteristics and learning capabilities. That is, AI has a certain ability to adaptively adjust parameters or update optimization models with changes in the environment, data, or tasks.

{\bf Scalability:} Over time, the computer hardware and software infrastructure for AI technology have improved, and the scale of AI models has grown accordingly. AI has been solving increasingly complex applications with increasing precision. With the development of new learning algorithms and architectures developed for deep neural networks (DNNs), AI is bound to have broader application prospects.

\subsubsection{Categories of AI} \label{subsec:CategoriesArtificialIntelligence}
As depicted in Fig.~\ref{fig:ClassificationArtificialIntelligence}, we first introduce the classification of AI. Then, we focus on several important branches of ML in the AI field. According to the classification of learning methods, ML can be divided into: supervised learning, unsupervised learning, and RL. We also discuss some typical network architecture in DL. 

{\bf Supervised Learning:} Supervised learning refers to training on labeled data to predict the type or value of new data \cite{goodfellow2016deep}. According to the different prediction results, supervised learning can be divided into two categories: classification and regression. In training, an objective function is usually given to measure the error (or distance) between the output and ground truth, and then its internal adjustable weights are modified to reduce the error via gradient descent. To improve the convergence speed and reduce the computational complexity, the stochastic gradient descent (SGD) \cite{bottou2007tradeoffs} method is often used in practice. SGD randomly selects a sample to compute the loss and gradient each time. Compared with more complex optimization techniques, this simple process of SGD often finds a good set of weights quickly. The common methods of supervised learning are K-nearest neighbor (KNN), decision tree, and logistic/linear regression. 

\begin{table*}[!t] 
	\centering
	\caption {Comparison of typical ML algorithms.} 
	\begin{tabular}{c|c|c}
		\hline \hline
		\tabincell{c}{\bf Learning Algorithms} &
		\tabincell{c}{\bf Characteristics} &
		\tabincell{c}{\bf Typical Methods}          
		\\ \hline 
		\bf Supervised Learning                          
		&
		\tabincell{c}{Predict the type or value of new data by training with labeled data.} 
		& 
		\tabincell{c}{KNN, Decision tree, \\ Logistic/Linear regression} 
		\\ \hline
		\bf Unsupervised Learning
		& 
		\tabincell{c}{Do data mining when the data has no labels.}
		& 
		\tabincell{c}{K-means clustering, \\ Principal component analysis}
		\\ \hline
		\bf Reinforcement Learning
		& 
		\tabincell{c}{A model consists of an Agent, which interacts with the environment. \\ The optimal policy is learned through a trial-and-error mechanism to \\ maximize long-term cumulative returns.} 
		& 
		DRL 
		\\ \hline \hline
	\end{tabular}
	\label{tab:ComparisonMLAlgorithms}
\end{table*}

{\bf Unsupervised Learning:} Unsupervised learning \cite{hinton2006fast} is to do data mining without labels. One of the important functions of unsupervised learning reflects in clustering, which is simply to classify data according to different features without labels. Typical methods of unsupervised learning include K-means clustering and principal component analysis, etc. An important premise of K-means clustering is that the difference between data can be measured by Euclidean distance. If it cannot be measured, it needs to be converted into a usable Euclidean distance measure. Principal component analysis is a statistical method. By using orthogonal transformation, the variables with correlation are changed into variables without correlation. The transformed variables are called principal components. The basic idea is to replace the initially correlated indicators with a set of independent comprehensive indicators.

{\bf Reinforcement Learning:} RL \cite{lecun2015deep,xue2022beam} is about obtaining rewards by interacting with the environment. Moreover, RL judges the quality of actions by the level of rewards and then learns the optimal strategy. Agent perceives the state information in the environment, searches for strategies, and selects the optimal action. This causes a state change and a return value to update the evaluation function. After completing a learning process, enter the next round of learning and training. The learning process is repeated cyclically and iteratively, until the conditions of stop rule are met, and then the learning is terminated. For large-scale station-action pair, deep reinforcement learning (DRL) \cite{arulkumaran2017deep} is an end-to-end perception and control system with strong generality. The principle framework of DRL is represented in Fig.~\ref{fig:ClassificationArtificialIntelligence}(d). The DRL learning process can be described as: (1) At each moment, the agent interacts with the environment to get a high-dimensional state, and uses DL to perceive the state to obtain a specific state feature representation; (2) The agent evaluates the value function of each action based on the expected return, and maps the current state to the corresponding action through a certain policy; (3) The environment reacts to this action and gets the next state. Through the continuous cycle of the above process, the optimal policy to achieve the goal can be finally obtained.

We summarize the characteristics and typical structures of several algorithms of ML discussed above as shown in the following TABLE~\ref{tab:ComparisonMLAlgorithms}.

{\bf Basic Network Architecture of DL:} The basic network structure is the convolutional neural network (CNN) \cite{lecun1989handwritten,lecun1998gradient}, which consists of an input layer, multiple convolutional layers, multiple pooling layers, a fully connected layer, and an output layer, as shown in Fig.~\ref{fig:ClassificationArtificialIntelligence}(a). The convolutional layer and the pooling layer are set alternately. In the convolutional layer, each neuron of the convolutional kernel is locally connected to its input, and weighted and summed with the local input through the corresponding connection weight. Then, the bias value is added to get the output value of the neuron. Because this process is equivalent to the convolution process, it is called CNN. CNN is easier to train and popularize than the fully connected network between adjacent layers, and is widely adopted in the field of computer vision.

The recurrent neural network (RNN) \cite{lipton2015critical} can process one element of the input sequence at a time. As shown in Fig.~\ref{fig:ClassificationArtificialIntelligence}(b), the RNN maintains a ``state vector" in its hidden units, which implicitly contains historical information for all past elements in the sequence. The output depends not only on the current input, but also on the information available in past moments or information available in future moments. With such special structure, RNNs are capable of providing memory for neural networks. 

Additionally, the generative adversarial network (GAN) \cite{goodfellow2014generative} is also a typical DL network that aims to learn a model capable of generating fake samples on real-distributed datasets. The basic structure of GAN is shown in Fig.~\ref{fig:ClassificationArtificialIntelligence}(c), which includes a generator G and a discriminator D. Both the generator and the discriminator can be implemented by DL networks. The discriminator is used to distinguish the fake samples generated by the generator from the real samples of the actual dataset. The task of the generator is to generate sample data such that the discriminator cannot distinguish between real samples and fake samples. When the generator produces samples that the discriminator cannot distinguish from the real samples, training is balanced. The applications of GAN in basic fields such as image generation, image translation, and speech images are very rich.

\subsubsection{AI for Wireless Communications} \label{subsec:AI6GNetworks}
In this section, we mainly describe the latest research progress of AI applied to 6G wireless communications. The combination of AI and 6G networks mainly contains in physical layer and upper layer.

{\bf AI for Physical Layer:} The main contents involve channel estimation \cite{kang2018deep,huang2018deep,he2018deep,borgerding2017amp,neumann2018learning}, signal detection \cite{ye2017power,samuel2017deep,he2018model,liao2020model,cao2021adaptive}, channel state information (CSI) feedback and reconstruction \cite{wen2018deep,wang2018deep,liu2019exploiting,guo2020compression,sangdeh2020lb}, channel decoding \cite{gruber2017deep,cammerer2017scaling,liang2018iterative,nachmani2016learning,nachmani2018deep}, and end-to-end wireless communications \cite{dorner2017deep,ye2018channel,damrath2016low,farsad2018sliding,mohamed2019model}. In massive multiple-input multiple-output (MIMO) beam mmWave scenarios, channel estimation is extremely challenging, especially in scenarios where antenna arrays are dense and receivers are equipped with limited radio frequency (RF) links. The work of \cite{kang2018deep} pioneered channel estimation by using the DL-based method in wireless energy transfer systems. In \cite{kang2018deep}, the authors developed an autoencoder-based channel estimation scheme, where the encoder is used to design pilots and the decoder is utilized to estimate the channel. The authors of \cite{huang2018deep} proposed a DL-based super-resolution channel estimation scheme in millimeter-wave massive MIMO systems. This scheme utilizes DNN for beam direction-of-arrival estimation. In contrast, there are some other DL-based channel estimation schemes that combine traditional algorithms with certain performance guarantees with DL algorithms. Reference \cite{he2018deep} designed a learned denoising-based approximate message passing (LDAMP) network. The LDAMP network takes the channel matrix as a two-dimensional image as input, and integrates denoising CNN into the iterative signal reconstruction algorithm for channel estimation. To improve the performance of sparse signal recovery, the authors of \cite{borgerding2017amp} proposed a learned approximate message passing (LAMP) network. LAMP directly expands the iterations of the AMP algorithm into the corresponding hierarchical network structure, whose linear transformation coefficients and nonlinear shrinkage parameters are jointly optimized by DNN. Furthermore, starting from the basic structure of the minimum mean square error (MMSE) algorithm, the work of \cite{neumann2018learning} developed a DL-based channel estimator, in which the estimated channel vector consists of conditional Gaussian random variables with random covariance matrices. To reduce the complexity of channel estimation, an MMSE-based CNN network is proposed to compensate for the error in \cite{neumann2018learning}. 

The authors in \cite{ye2017power} applied DNN to tackle the problem of signal detection in orthogonal frequency division multiplexing (FDM) systems. Different from traditional wireless communication, \cite{ye2017power} regarded channel estimation and signal detection as a whole, and directly uses DNN to realize the mapping from the received signal to the original signal bits. The work in \cite{samuel2017deep} investigated the signal reconstruction problem of the MIMO system, and proposed a signal detection algorithm entitled Detection Network (DetNet). DetNet is based on the maximum likelihood method by adding the gradient descent algorithm to generate a DL network. Based on the orthogonal AMP (OAMP) iterative algorithm combined with the DL network, OAMP-Net was proposed in \cite{he2018model}. The purpose of OAMP-Net is to add adjustable training parameters on the basis of the original algorithm to further improve the signal detection performance of the existing algorithm. With the advantage of fewer trainable parameters, the model-driven detection network \cite{liao2020model} was designed to improve detection performance by expanding a specific iterative detector and adding some trainable parameters. In addition, an adaptive signal detection method named JC-Net for massive MIMO systems was proposed in \cite{cao2021adaptive}. JC-Net has a foundation for the traditional Jacobi detector, adding trainable parameters to improve the convergence speed and perform corresponding soft projection. In FDM networks, the BS requires to attain downlink CSI feedback to perform precoding and achieve performance gains. However, there are many configured antennas in massive MIMO systems, so the feedback overhead of the complete CSI becomes extremely huge. The work of \cite{wen2018deep} presented a CNN-based CSI perception and recovery mechanism named CsiNet. Since then, DL-based CSI compression techniques have attracted a lot of attention \cite{wang2018deep,liu2019exploiting,guo2020compression,sangdeh2020lb}. On the basis of \cite{wen2018deep}, the authors of \cite{wang2018deep} provided a real-time long short-term memory (LSTM)-based CSI feedback architecture entitled CsiNet-LSTM, which employs temporal correlation to improve the feedback accuracy of time-varying channels. CsiNet-LSTM can accomplish a trade-off between compression ratio, CSI reconstruction quality, and complexity. On the basis of the high correlation of amplitudes between bidirectional channels in the delay domain, DualNet was proposed in \cite{liu2019exploiting} to use uplink amplitude information to help reconstruct downlink channel amplitudes. The CSI feedback and reconstruction algorithms in \cite{wen2018deep,wang2018deep,liu2019exploiting} rely on a large amount of data for offline training, and the network complexity is high. The work of \cite{guo2020compression} focused on the complexity of the neural network. The experimental results in \cite{sangdeh2020lb} demonstrated that the DL-based channel feedback framework can reduce the air time overhead by an average of 73\% and improve the throughput by about 69\% compared with the 802.11 feedback protocol.   

The authors in \cite{gruber2017deep} developed a DNN-based channel decoding method. This paper draws two conclusions about the application of DL to channel decoding: 1) Structured codes such as polar codes are easier to learn than random codes; 2) For structured codes, DL networks can decode untrained codewords. However, this proposed method is neither suitable for random codes nor codewords with long code lengths, and has great limitations. On the basis of the traditional polar code iterative decoding algorithm, the work of \cite{cammerer2017scaling} presented a DL polar code decoding network that separates sub-blocks. The decoding algorithm in \cite{cammerer2017scaling} is a highly parallel decoding algorithm. Compared with the decoding algorithm in \cite{gruber2017deep}, the algorithm of \cite{cammerer2017scaling} significantly reduced the number of training times and the complexity of the network structure under the condition of comparable performance. Reference \cite{liang2018iterative} conducted an iterative channel decoding algorithm: {belief propagation (BP)-CNN}. The algorithm concatenates the CNN with the standard BP decoder to estimate information bits in a noisy environment. For high density parity check (HDPC) codes, the performance of the BP algorithm is relatively poor. Nachmani et al. successively proposed the BP-DNN algorithm \cite{nachmani2016learning} and the BP-RNN algorithm \cite{nachmani2018deep}, which combined the DNN and RNN networks with BP algorithms to improve the performance of BP algorithms applied to HDPC. Reference \cite{dorner2017deep} put forward an end-to-end wireless communication system model, which explains the feasibility of replacing the processing module of the physical layer by DNN. The authors of \cite{ye2018channel} provided a differentiable channel computational model, which can be used for supervised autoencoder training. Since then, many non-modeled methods \cite{damrath2016low,farsad2018sliding} have been developed based on synchronization-around methods, none of which require any channel knowledge and can be directly executed on real hardware. In \cite{mohamed2019model}, the authors treated the communication system as an end-to-end DRL autoencoder. This technique does not require any information about the actual channel model.  

\begin{figure*}[!t]
	\centering    
	\includegraphics[width=0.82\linewidth]{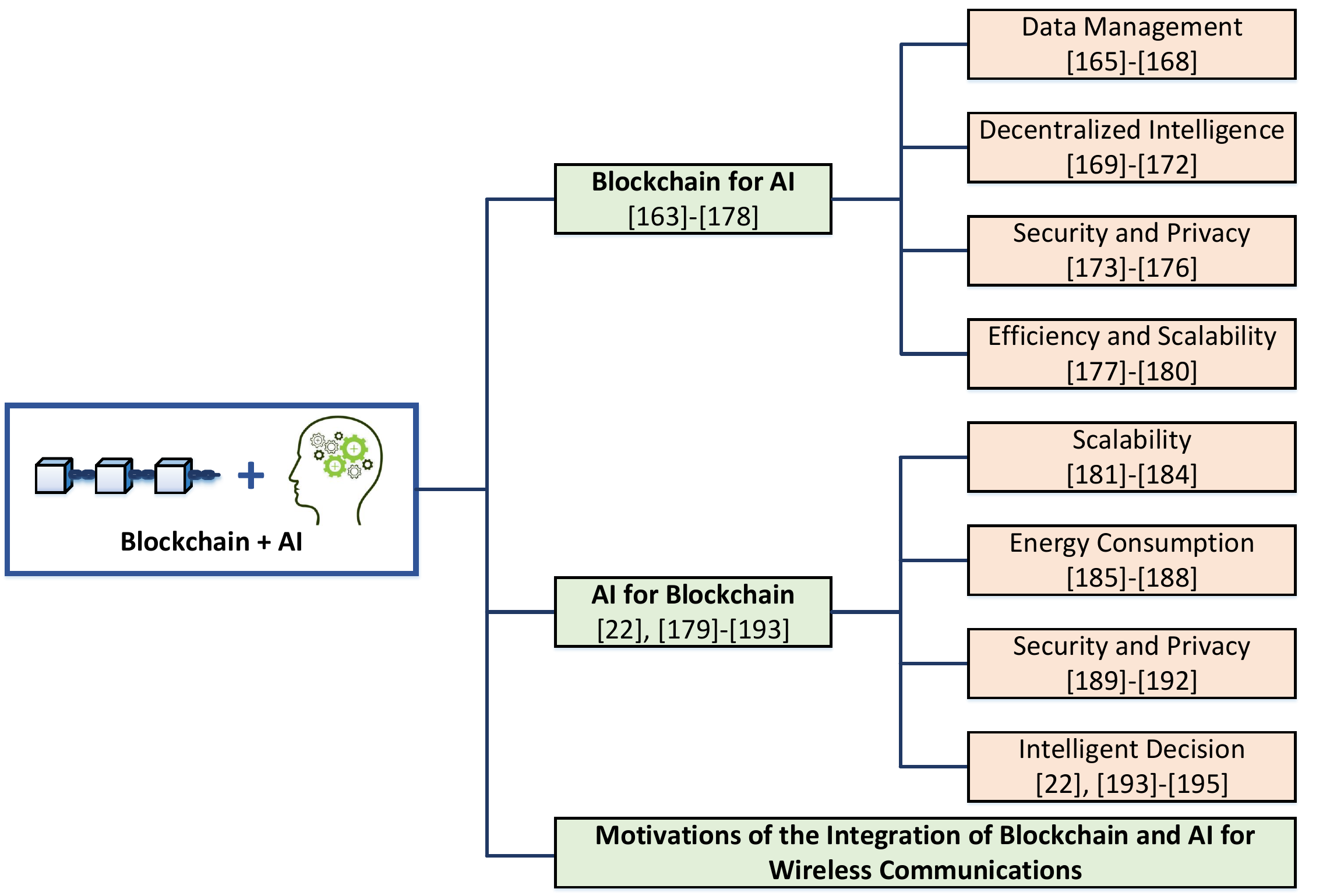}   
	\caption{{{Taxonomy of the integration of blockchain and AI.}}}
	\label{fig:TaxonomyBlockchainAI}
\end{figure*}

{\bf AI for Upper Layer:} In recent years, AI has been introduced into the upper layers of wireless communications to tackle various problems, thereby enabling near-optimal network performance. For example, since artificial neural networks have the approximation characteristics of universal functions, \cite{zhao2020power} and \cite{d2019uplink} adopted a data-driven approach to allow the training model to autonomously learn user access and power allocation strategies. Supervised learning requires predicting the labels of the training data, resulting in excessive data preprocessing burden. So, from the perspective of unsupervised learning, the authors of \cite{rajapaksha2021deep} leveraged a feed-forward neural network to autonomously learn the optimal power allocation. Bypassing channel estimation, the work in \cite{cui2019spatial} efficiently scheduled interfering links based only on the geographic locations of transmitters and receivers via DL algorithms. In multi-cell systems, reference \cite{xu2019realtime} approximated optimal link scheduling and power control through DNNs.  Specifically, a matching link schedule was estimated using the deep Q-network, and then power was allocated to the corresponding link schedule.

As a data-driven ML method, DRL can directly learn dynamic environmental laws and obtain optimal decisions. Therefore, DRL can endow the network with the ability to self-optimize management according to the dynamic environment, making intelligent communication possible. Next, we focus on the application of DRL in the upper layer of wireless communications. For example, reference \cite{li2018intelligent} considered a power control problem in cognitive radio. Here, to improve the spectrum usage rate, secondary users performed communication by occupying the spectrum of primary users. To satisfy the service quality of primary and secondary users, the author of \cite{li2018intelligent} proposed a DRL-enabled power control scheme. Reference \cite{li2018intelligent} was aimed at the single-user power control problem, which cannot be applied to multi-user scenarios. To this end, the work of \cite{nasir2019multi} discussed the problem of multi-user power resource allocation in cellular networks, where the goal was to maximize the weighted sum-rate of the entire network. Moreover, \cite{tan2020deep} extended \cite{nasir2019multi} to multi-user {device-to-device (D2D)} communication scenarios. The authors of \cite{sadeghi2019deep} used DRL to analyze the user's data request, and replaced the file in the cache as per the user's request rule. A DRL-empowered computing resource allocation scheme was presented in \cite{ren2019federated}. In this scheme, IoT devices adopted DRL algorithms to determine the power of each computing task to be executed locally, and a power of 0 meant that the computing task is executed in MEC servers. In addition, references \cite{wang2019edge,he2018trust,he2017integrated,ndikumana2019joint} also successfully employed DRL algorithms in the joint optimization problem of caching and computing resources, indicating that DRL has a strong application prospect in managing network resources.

In \cite{xu2019load}, DRL was used to realize intelligent horizontal handover between BSs. The work of \cite{zhao2019deep} further attempted to combine access control and resource allocation, and considered the DRL algorithm to solve the joint optimization problem of user access and channel allocation in multi-layer BS cellular networks. As the size of the network escalates, the probability of network failure also increases. The authors of \cite{mismar2018deep} attempted to apply DRL to network fault self-healing. To enhance energy efficiency and reduce costs, wireless networks need to dynamically turn BSs on and off according to user traffic demands. In view of the dynamic randomness of traffic demands, \cite{ye2019drag} proposed to apply DL to analyze and predict the traffic of each BS, and then used DRL to control the switch of BSs according to the predicted traffic. In addition to \cite{ye2019drag}, the work of \cite{liu2018deepnap} also introduced a DRL-supported intelligent sleep strategy for BSs to reduce network energy consumption. In sparsely populated areas, UAVs can be leveraged as air BSs to serve terrestrial communication terminals. Considering the coverage limitation of the UAV and the moving variation of the air-to-ground channel, the authors in \cite{wu20193d} discussed to use DRL algorithms with deep Q-learning for deployment planning of air BSs.

\subsection{{Integration of Blockchain and AI}} \label{sec:IntergrationBlockchainAI}
In recent years, the frontier technologies of blockchain and AI have aroused widespread attention and in-depth research in academia and industry. Blockchain technology has the characteristics of decentralization, anonymity, openness and transparency, and immutability. However, the blockchain needs to be improved urgently in terms of scalability, energy consumption, and security. As a powerful analysis and decision-making tool, AI can predict and analyze data in real-time scenarios and make optimal decisions. Nevertheless, the centralized structure of AI and its demand for security and credibility have greatly limited the wide application of AI. Therefore, there is complementary potential for the combination of blockchain and AI. As shown in Fig.~\ref{fig:TaxonomyBlockchainAI}, we respectively elaborate and analyze from the two aspects of blockchain for AI and AI for blockchain.

\subsubsection{Blockchain for AI} \label{subsec:BlockchainforAI}
From the perspective of AI technology, blockchain as a trusted platform can create a secure, immutable, and distributed system for AI. In this secure system, no third-party participation and management are required, users trust each other and share data. Accordingly, based on the huge and reliable data set, the accuracy of the agent's decision-making is improved. We introduce blockchain-driven AI from some aspects below, including data management \cite{harris2019decentralized,zhang2018blockchain,wang2019securing,liu2019blockchainenabled}, decentralized intelligence \cite{mendis2021ablockchain,wang2020aiat,korkmaz2020chainfl,lin2019making}, security and privacy \cite{gupta2021when,weng2021deepchain,shayan2021biscotti,feng2021blockchainbased}, and efficiency and scalability \cite{nassar2020blockchain,sarpatwar2019towards,kang2019incentive,bao2019flchain}.

{\bf Data Management:} The massive amount of AI data lacks a consolidated and efficient sharing mechanism and management method. The poor maintainability of open-source data sets leads to uneven data quality, and the data is not centralized and unified. The distributed database of blockchain efficiently collects, shares, and stores the data of each node, so that every participant on the network can access the data. This can provide AI with broader data access and more efficient data monetization mechanisms. For instance, to make the updating of AI models more efficient, based on blockchain technology, the authors of \cite{harris2019decentralized} proposed the novel configurable distributed AI framework, where participants collaborated to construct datasets and used smart contracts to host continuously updated AI models. In order to break the data barriers between different mobile operators, \cite{zhang2018blockchain} designed a blockchain-empowered data sharing framework and a Hyperledger-based prototype system. This system utilized smart contract-based monitoring and fine-grained data access control to create a safe and reliable environment for data sharing. Using blockchain to help AI manage trusted data, the work in \cite{wang2019securing} introduced a secure large-scale Internet architecture called SecNet. The SecNet can realize secure data storage, computing, and sharing, and enhance AI with a large number of data sources. The work in \cite{liu2019blockchainenabled} demonstrated a blockchain-enabled joint framework for efficient data acquisition and secure sharing. This proposed framework used DRL to achieve the maximum amount of collected data, and leveraged the Ethereum blockchain technology to ensure the security and reliability of data sharing.

{\bf Decentralized Intelligence:} Through AI algorithms, learning results and models can be obtained from massive data. Due to the distribution of IoT devices or edge computing devices and the data heterogeneity, the cooperation of multiple devices is required to complete complex model training tasks. That is, different devices need to share data for data analysis and prediction. Local learning models can also be shared across devices and then aggregated. Blockchain technology can guarantee that AI completes the interaction of data or models between devices in the decentralized environment. To realize asynchronous cooperative computing among untrusted nodes, the work in \cite{mendis2021ablockchain} developed a decentralized, privacy-preserving, and secure computing paradigm, which adopted various technical means such as blockchain, decentralized learning, and homomorphic encryption. A blockchain-assisted distributed secure multi-party learning architecture was proposed in \cite{wang2020aiat}. Specifically, the authors formulated two types of Byzantine attacks, as well as elaborated ``off-chain" and ``on-chain" mining schemes. In response to the single point of failure problem of federated learning (FL), the work in \cite{korkmaz2020chainfl} provided a decentralized FL scheme entitled ChainFL. The ChainFL utilized blockchain to delegate the responsibility of storing and aggregating models to nodes on the network without requiring any central server. In edge AI-supported IoT networks, to break knowledge silos, the authors of \cite{lin2019making} proposed a P2P knowledge payment sharing architecture, which made use of the knowledge consortium blockchain to ensure that knowledge management and market transactions are safe and efficient. This knowledge consortium blockchain included the new encrypted currency knowledge currency, smart contracts, and new transaction consensus mechanism proof.

{\bf Security and Privacy:} For AI technology, the greater amount of having data, the higher accuracy of its training model. However, if a small part of this data has security issues, the validity of the data will affect the system's decision-making  accordingly and thus the overall performance of the system. Fortunately, blockchain has many technologies such as anonymity, immutability, interface access control, and signature authentication and authorization to ensure the security and privacy of transaction data, and to provide quality assurance for the data required for AI model training. The work in \cite{gupta2021when} proposed a blockchain-authorized edge intelligence system, which assured the security, privacy, latency, and efficiency of edge device data. Here, the public blockchain guaranteed the security and privacy of data of edge devices, while the private blockchain ensured secure communication between edge intelligent servers. The authors of \cite{weng2021deepchain} demonstrated a distributed DL architecture named DeepChain. DeepChain utilized the value-driven incentive mechanism of blockchain to encourage parties to collaborate in DL model training and share the obtained local gradients. Meanwhile, DeepChain guaranteed the privacy of local gradients for each participant and provided auditability for the entire training process. To prevent malicious attacks on AI models, the work in \cite{shayan2021biscotti} introduced Biscotti, which was a fully decentralized P2P large-scale multi-party learning scheme. The Biscotti adopted blockchain and cryptographic primitives to coordinate the privacy-preserving ML process among peer nodes. The authors of \cite{feng2021blockchainbased} presented the blockchain-assisted asynchronous FL (BAFL) architecture, where the blockchain ensured that model data cannot be tampered with, and assured decentralized and secure data storage, and asynchronous learning accelerates global aggregation. The proposed BAFL guaranteed that each device uploaded the local model whenever the global aggregation can converge the global model faster.

{\bf Efficiency and Scalability:} When using AI techniques, such as DL, it is difficult for people to understand what is in the black box and explain the decisions made by AI systems, so AI cannot be verified or trusted. Furthermore, without appropriate incentive mechanisms, various parties may be reluctant to participate in data training. The above-mentioned problems will reduce the efficiency and scalability of AI systems. Blockchain can track every link in the data processing and decision-making chain for explainable AI. Appropriate incentive mechanisms can also be introduced from blockchain. The transparent and cost-effective incentive mechanism design can be implemented, which will greatly improve the enthusiasm of all parties in AI systems to take part in the training. A blockchain-based framework for more trustworthy and explainable AI was presented in \cite{nassar2020blockchain}. This framework leveraged smart contracts to record and manage interactions, as well as provided consensus for trusted oracles. The proposed framework also addressed decentralized storage, registry, and reputation supporting services. To obtain better trusted AI, the authors of \cite{sarpatwar2019towards} designed a blockchain-enabled FL system, which used the blockchain to track the source information of the trained model. Aiming at the incentive mechanism problem of FL, the work in \cite{kang2019incentive} presented a reputation-based miner selection scheme, and designed an efficient incentive mechanism by adopting a multi-weight subjective logic model to evaluate the user's credit. The proposed scheme also leveraged the consortium blockchain to achieve secure reputation management for miners in a decentralized manner. In \cite{bao2019flchain}, the authors described the FLChain framework based on trust and incentive. The FLChain saved miners' information and verifiable training details for public audits. The incentive mechanism of the FLChain encouraged honest and trustworthy miners. Otherwise, malicious nodes will be punished, so as to maintain a healthy and reliable public platform.

\subsubsection{AI for Blockchain} \label{subsec:AIforBlockchain}
From the perspective of blockchain technology, its scalability and system energy consumption can be optimized through AI algorithms. Using AI algorithms in blockchain networks, security vulnerabilities brought about by the implementation of smart contracts and consensus mechanisms can be identified and detected. We will represent AI-driven blockchain from four aspects below, including scalability \cite{liu2019performance,yun2021dqnbased,qiu2021deepreinforcement,zhang2020skychain}, energy consumption \cite{bravo2019proof,chenli2019energy,baldominos2019coin,chen2018ai}, security and privacy \cite{chen2018detecting,salimitari2019ai,liang2021data,scicchitano2020deep}, and intelligent decision \cite{wang2021blockchain,mudassir2020time,singh2019prediction,mohammadiprediction}.

{\bf Scalability:} Currently, as the number of transactions increasing significantly, scalability is the biggest barrier to the widespread application of blockchain technology. In blockchain systems, the core of scalability is to tackle the problems of transaction throughput and transaction speed. Due to the characteristics of decentralization and network-wide broadcasting, each node on the blockchain will record transactions generated by the entire network, leading to low efficiency. AI can introduce DRL or data sharding technology to propose new solutions to blockchain scalability issues and improve system efficiency. In \cite{liu2019performance}, through the DRL algorithm, the agent dynamically selected different consensus algorithms and block production nodes, and adjusted the block size and time interval. The agent found optimal parameters to improve the scalability, while ensuring the decentralization, latency, and security of blockchain networks. To obtain better throughput, the authors of \cite{yun2021dqnbased} adopted the deep Q network algorithm to dynamically adjust the block size, time interval, and the number of shards to seek the optimal related parameters, while meeting the security of the system. The work in \cite{qiu2021deepreinforcement} studied a DRL-enabled adaptive blockchain scheme, which improved scalability and met the needs of different users. Specifically, according to the service quality requirements of users, the DRL algorithm selected the most suitable consensus protocol for blockchain systems. To overcome the limitations of existing blockchain static sharding, \cite{zhang2020skychain } introduced a DRL-based dynamic sharding blockchain framework called SkyChain. In the dynamic environment of blockchain systems, this presented SkyChain can dynamically adjust the resharding interval, number of shards, and block size in order to maintain a long-term balance between performance and security.

{\bf Energy Consumption:} Blockchain mining requires a large amount of computing power and electricity resources. At present, Bitcoin consumes about 2.55 billion watts of electricity every year, almost the same as the annual electricity consumption of some small countries.  If the energy consumption problem cannot be solved well, the value of the blockchain itself will be diluted. To avoid excessive consumption of computing resources and energy resources in this mining process, AI algorithms can understand the blockchain network process and architecture, and explore a more effective consensus mechanism, which can make transactions on blockchain networks execute faster.

A Proof-of-Learning consensus protocol was formed by combining ML algorithms in \cite{bravo2019proof}. This Proof-of-Learning protocol performed model training through ML of given tasks. Then, the ranking was based on the minimum loss function value. Finally, the optimal model parameters were selected and verified by other mining nodes to achieve distributed consensus. Through combining DL algorithms, the work of \cite{chenli2019energy} proposed a Proof-of-Deep-Learning consensus protocol, which forced the agent to conduct DL model training, and proposed the training model as a proof of effectiveness. Only when an appropriate DL model was generated, can miners reach the consensus and generate new blocks. In response to the energy consumption problem in blockchain networks, the authors of \cite{baldominos2019coin} introduced a Proof-of-Useful-Work energy-saving consensus protocol. The proposed protocol required training a DL model during the mining process, and mining new blocks only when the performance of the training model exceeded a given threshold. The work in \cite{chen2018ai} demonstrated an AI-enabled node selection algorithm that exploited the nearly complementary information of each node and relied on a specially designed CNN to reach consensus. In order to ensure the decentralization and security of the network, dynamic thresholds were used to obtain super nodes and random nodes.

{\bf Security and Privacy:} The decentralized power of blockchain may be at risk of abuse, especially since the smart contracts and consensus mechanisms in blockchain technology are vulnerable to malicious network attacks or tampering. As more and more personal data is stored in blockchain systems, data privacy protection becomes critical. We employ AI-assisted methods to identify and detect security vulnerabilities, greatly improving the security and privacy of blockchains. As an illustration, the work of \cite{chen2018detecting} learned by extracting relevant features from user accounts and operation codes of a large number of smart contracts, and used the ML algorithm XGBoost to detect whether there is a potential Ponzi scheme in the smart contract. Ponzi scheme is a classic investment fraud, and it also has a blockchain-based form. The essence of Ponzi is that the investment of new investors is the return of old investors. On the Hyperledger Fabric blockchain platform and with a relatively low tolerance for malicious activities, the authors of \cite{salimitari2019ai} designed an external detection algorithm based on supervised ML before the consensus protocol as the consensus of the previous step. This proposed detection algorithm verified the new data compatibility, and discarded suspicious data to improve the network's fault tolerance for the second-step consensus. In blockchain-based systems, a data fusion scheme based on collaborative clustering features was represented in \cite{liang2021data}. The data fusion scheme applied AI algorithms to train and analyze data clusters to detect abnormal intrusion behaviors. To effectively detect abnormal behavior of blockchain systems, the work in \cite{scicchitano2020deep} proposed an encoder-decoder-based DL model, which was an unsupervised model trained with aggregated information extracted by monitoring blockchain transactions.

\begin{figure*}[!t]
	\centering    
	\includegraphics[width=0.9\linewidth]{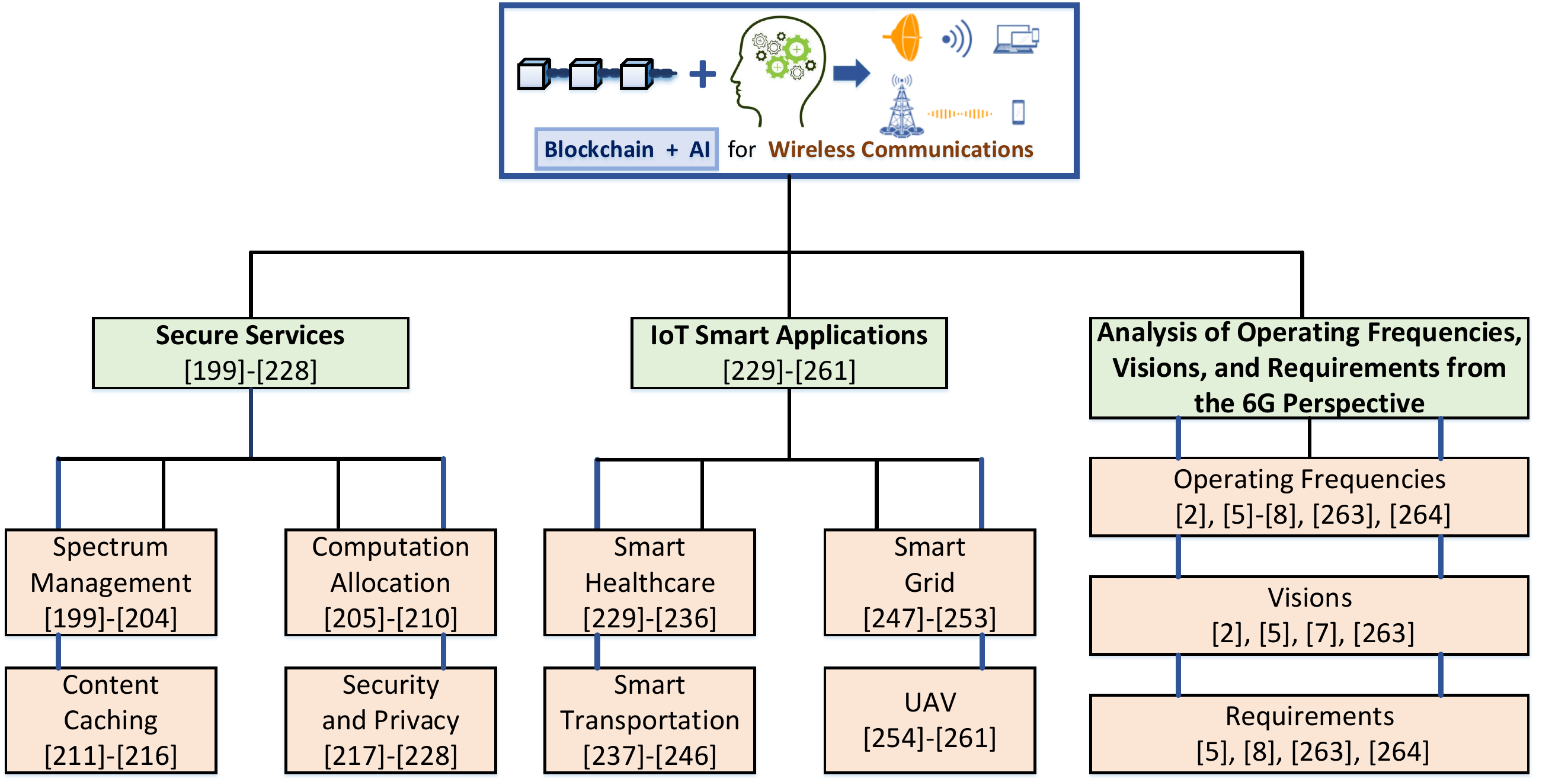}   
	\caption{{{Taxonomy of the integration of blockchain and AI for wireless communications.}}}
	\label{fig:TaxonomyBlockchainAI6G}
\end{figure*}

{\bf Intelligent Decision:} With the rise of blockchain technology, more people turn to study blockchain-empowered application projects. Investors want to predict some important behaviors of blockchain systems, such as cryptocurrency prices, transaction confirmation times, and blockchain forks. We consider embedding AI algorithms into blockchain systems and allowing AI to optimize or make decisions for the entire system, which is more conducive to investors making correct policies. In \cite{wang2021blockchain}, without knowing the details of the blockchain network model, a multi-dimensional RL algorithm was proposed to solve the mining problem with the Markov decision process. The designed algorithm can obtain the near-optimal mining strategy solution in the time-varying blockchain network. In both short-term (1 day and 7 days) and medium-term (30 days and 90 days) time periods, the authors of \cite{mudassir2020time} employed classification and regression models of ML to predict the trend of Bitcoin price. The results showed that the proposed four types of ML models predicted the actual Bitcoin price with a very low error rate. Applying ML, the work in \cite{singh2019prediction} demonstrated an Ethereum prediction model, which can predict transaction execution times in Ethereum systems. The transaction execution time refers to the time frame within which a miner node accepts and includes a transaction in a block. To reduce the huge risk and cost brought by the fork, the work of \cite{mohammadiprediction} adopted an ML method to predict the blockchain fork, and compared the prediction accuracy of the fork by four well-known ML methods, namely K Near Neighbor, Naive Bayes, Decision Tree, and Multilayer Perceptron.

\subsubsection{Motivations of the Integration of Blockchain and AI for Wireless Communications} \label{subsec:MotivationIntergrationBlockchainAI6G}
Blockchain can establish a secure and decentralized resource sharing environment. AI can solve some problems with uncertain, time-varying, and complex characteristics. {As shown in TABLE~\ref{tab:ComparisonSolutions}, we summarize and compare blockchain for 5G/6G, AI for 5G/6G, blockchain for AI, and AI for blockchain. We conduct a critical and original discussion of these existing solutions, highlighting the advantages, disadvantages, and main findings of various solutions.} Although blockchain and AI are promising technologies to be applied in 6G networks, there are still many challenges and unresolved problems. Both blockchain and AI have attracted significant attention recently. The combination of these two technologies may further improve the performance of 6G networks. In the first place, to more systematically understand the integration and application of blockchain and AI technologies for 6G networks, we summarize the benefits of blockchain for AI and the benefits of AI for blockchain, respectively. For details, please refer to the above subsections: Section~\ref{subsec:BlockchainforAI} and Section~\ref{subsec:AIforBlockchain}. Then, we briefly describe the benefits that fusing blockchain and AI can bring to 6G networks.

\begin{table*}[!t] 
	\centering
	\caption {{Comparison of existing solutions.}} 
	\begin{tabular}{c|c|c|c}
		\hline \hline
		\tabincell{c}{\bf {Solution}} &
		\tabincell{c}{\bf {Advantages}} &
		\tabincell{c}{\bf {Disadvantages}} &
		\tabincell{c}{\bf {Main Findings}} 
		\\ \hline 
		{\tabincell{c}{\bf Blockchain \\ \bf for 5G/6G \\ \bf [57]-[97]}}                         
		&
		{\tabincell{l}{The integration of blockchain and \\ 6G will provide a strong security \\ guarantee for the construction of \\ a safe and credible communication \\ ecosystem.}}
		& 
		{\tabincell{l}{When there are massive blockchain \\ applications and nodes communicat- \\ ing with each other, the 6G network \\ may face uncertain local network \\ congestion.}}
		&
		{\tabincell{l}{While blockchain poses challenges to \\  the stability of 6G networks, it can \\ also provide protection for the security \\ of 6G networks and increase the value \\ of data.}}  
		\\ \hline
		{\tabincell{c}{\bf AI for \\ \bf 5G/6G \\ \bf [112]-[156]}}
		& 
		{\tabincell{l}{AI enhances the performance of \\ a specific module of 6G systems, \\ including improving accuracy and \\ reducing complexity, and also int- \\ egrates multiple communication \\ modules to break the existing mo- \\ dular communication architecture.}} 
		& 
		{\tabincell{l}{Inefficient data management schemes \\ and high overhead of information \\ exchange among communication par- \\ ticipants are key bottlenecks. Data \\ security and privacy issues are rece- \\ iving increasing attention.}}
		&  
		{\tabincell{l}{6G network realizes interconnected \\ intelligence by supporting AI funct- \\ ions, and adopts a centralized network \\ architecture, which is vulnerable to \\ hacker attacks.}}
		\\ \hline
		{\tabincell{c}{\bf Blockchain \\ \bf for AI \\ \bf [157]-[172]}}
		& 
		{\tabincell{l}{Blockchain can conduct a secure, \\ immutable, and distributed syst- \\ em for AI. Users trust each other \\ and share data. The performance \\ of AI algorithms and decision- \\ making are effectively upgraded.}}
		& 
		{\tabincell{l}{The execution results of smart contracts \\ in blockchains are often deterministic. \\ While, the execution results of AI are \\ usually uncertain, random, and unpre- \\ dictable in most cases.}} 
		& 
		{\tabincell{l}{The contradiction between blockchain \\ and AI poses certain challenges for \\ AI embedded in blockchain to opti- \\ mize the execution decisions.}}
		\\ \hline
		{\tabincell{c}{\bf AI for \\ \bf Blockchain \\ \bf [173]-[187]}}
		& 
		{\tabincell{l}{AI can enhance the performance \\ of blockchains in terms of scalab- \\ility, energy consumption, securi- \\ ty and privacy, and intelligent \\ decision.}}
		& 
		{\tabincell{l}{With the explosive growth of data in \\ AI-assisted blockchain systems, the \\ massive unlabeled and unclassified \\ datasets are intractable for AI training.}}
		& 
		{\tabincell{l}{Using AI algorithms, the scalability and \\ energy consumption issues of blockchain \\ can be mitigated, but AI training also \\ faces challenges.}}
		\\ \hline \hline
	\end{tabular}
	\label{tab:ComparisonSolutions}
\end{table*} 

On the one hand, blockchain can improve AI in terms of data management, decentralized intelligence, security and privacy, and efficiency and scalability. Firstly, blockchain collects, shares, and stores data for AI, so that every participant on the network can access the data. Blockchain-supported methods can provide AI with more efficient data management mechanisms and wider data access. Secondly, blockchain can ensure that AI can complete the collaborative interaction of data or models between devices in a decentralized environment. In addition, blockchain uses its own anonymity, immutability, interface access control, signature authentication and authorization, and other technologies to safeguard the security and privacy of transaction data in the AI system. Finally, blockchain can track every link in the data processing and decision-making chain for explainable AI. Transparent and cost-effective incentive mechanisms can also be designed, which will effectively upgrade the efficiency and scalability of AI systems. 

On the other hand, AI can enhance the performance of blockchains in terms of scalability, energy consumption, security and privacy, and intelligent decision. First of all, AI can introduce DRL or data sharding technology, to propose new solutions for blockchain scalability issues and ameliorate system efficiency. Furthermore, in order to avoid the massive resource consumption of blockchain, AI algorithms can dissect the blockchain network process and architecture, as well as explore a more effective AI-based consensus mechanism based on AI, so that transactions on the blockchain can be executed faster. Next, AI-assisted methods are used to identify and detect security vulnerabilities, which remarkably augments the security and privacy of blockchains. Ultimately, we can consider embedding AI algorithms into blockchain systems and letting AI optimize or make decisions for the entire system, which is more conducive to investors making make correct decisions.

According to the above analysis, the amalgamation of blockchain and AI has complementary potential. Blockchain can conduct a secure, immutable, and distributed system for AI technologies. In this system, users trust each other and share data. Based on a huge and reliable data set, the performance of AI algorithms and decision-making can be effectively upgraded. Using AI algorithms, the scalability and energy consumption issues of blockchain can be mitigated, and its security vulnerabilities can also be identified and detected. The integration of AI and blockchain is not only to enhance each other, but also to push and optimize various services and applications for 6G scenarios in the process of mutual promotion. In this case, a reliable, secure, and ultra-low latency network environment can be provided for 6G wireless communications. Consequently, the research on the integration of blockchain and AI is extremely important and worth expecting in 6G networks.

{{In this section, we have provided a comprehensive overview of the fundamental concepts, characteristics, and categories of blockchain and AI. We have also discussed the classic applications of both technologies in wireless communication systems. Then, we systematically summarized the integration of blockchain and AI from two directions: blockchain-assisted AI and AI-assisted blockchain. Furthermore, we analyzed the advantages of integrating blockchain and AI for wireless communication systems. Through this section, we have gained valuable insights into the opportunities and challenges of leveraging blockchain and AI in wireless communication systems. We have also recognized the importance of considering different integration approaches and identifying suitable use cases to maximize the potential of these technologies. Overall, this section provides a solid foundation for further exploration and analysis of the integration of blockchain and AI in wireless communication systems.
}}

\section{Integration of Blockchain and AI for Wireless Communications}
\label{sec:IntegrationBlockchainAI6G}
\begin{figure*}[!t]
	\centering    
	\includegraphics[width=1.0\linewidth]{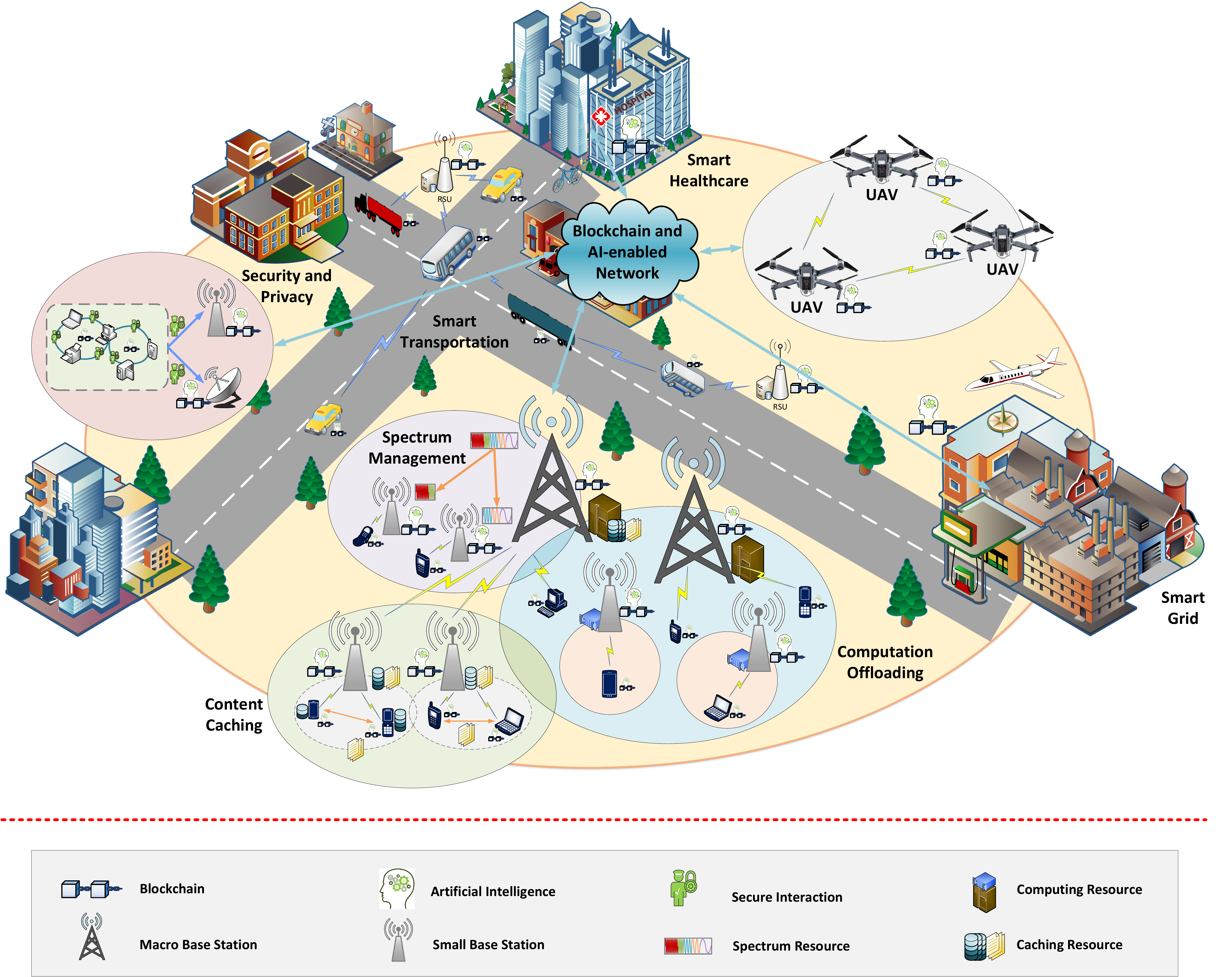}   
	\caption{The convergence of the integration of blockchain and AI for wireless communications.}
	\label{fig:ConvergenceBlockchainAI6G}
\end{figure*}

For the existing problems of blockchain and AI, the integration of these two technologies can complement each other. Facing the 6G era, the network will meet new application scenarios and new performance requirements. Diverse applications, communication scenarios, ultra-heterogeneous network connections, and service requirements for extreme performance all put forward higher demands on mobile communication networks \cite{liu2020blockchain}. The merger of blockchain and AI can not only play to their respective advantages\cite{pan2020blockchain,zhang2019edge}, but also better bring optimization and improvement to various services and applications in 6G networks \cite{jameel2020reinforcement,wu2021deep,rathore2021deep}. In this section, we will discuss broadly the applications of merging blockchain and AI in 6G networks, including 6G secure services \cite{hu2021blockchain,maksymyuk2022ai,lu2020blockchain,lu2020communication,guo2019adaptive,gur2020expansive,qiu2018blockchain,he2020blockchain,manogaran2021artificial,nguyen2021secure,liao2020blockchain,liao2021blockchain,qiu2020ai,qian2020blockchain,dai2020deep,li2020resource,cui2020creat,zhang2020deep,waheed2020security,dhieb2020scalable,wang2021enabling,tian2021blockchain,lin2020blockchain,kumar2021privacy,zhang2020ait,zhang2020blockchain,otoum2021securing,lu2019blockchain,zhang2021bc,liu2021blockchain} and 6G IoT smart applications\cite{bhattacharya2019bindaas,al2021reinforcement,veeramakali2021intelligent,kumar2021blockchain,otoum2021preventing,mallikarjuna2021blockchain,guo2021smartphone,gupta2020bats,al2021enabling,hassija2020traffic,tiba2020secure,song2020blockchain,zhang2019blockchain,jiang2020intelligent,fu2020autonomous,pokhrel2020federated,lu2020blockchainempowered,chai2020hierarchical,kumar2020distributed,keshk2019privacy,wang2020aebis,ferrag2019deepcoin,jamil2021peer,gao2021fogchain,li2021integration,singh2020deeplearning,feng2021blockchainempowered,asheralieva2019distributeddynamic,pokhrel2021blockchainbrings,islam2021blockchainbased,gumaei2021deeplearningand,aftab2020blockml,pokhrel2020federatedlearningmeets} as depicted in Fig.~\ref{fig:TaxonomyBlockchainAI6G}. The convergence of the integration of blockchain and AI for wireless communications is illustrated in Fig.~\ref{fig:ConvergenceBlockchainAI6G}. {{Furthermore, we thoroughly discuss operating frequencies, visions, and requirements from the 6G perspective.}}

\subsection{Secure Services} \label{subsec:6GSecureServices}
As mentioned in the previous section, the combination of blockchain and AI can not only promote each other, but also provide better services. In 6G networks, wireless resources such as spectrum, computing, and caching are some of the most concerned services. The development of 6G networks will bring explosive growth of user data, and security and privacy services are also the keys to improving the overall performance. In this subsection, we will focus on some key 6G secure services, where blockchain and AI are simultaneously applied, including spectrum management \cite{hu2021blockchain,maksymyuk2022ai,lu2020blockchain,lu2020communication,guo2019adaptive,gur2020expansive}, computation allocation \cite{qiu2018blockchain,he2020blockchain,manogaran2021artificial,nguyen2021secure,liao2020blockchain,liao2021blockchain}, content caching \cite{qiu2020ai,qian2020blockchain,dai2020deep,li2020resource,cui2020creat,zhang2020deep}, and security and privacy \cite{waheed2020security,dhieb2020scalable,wang2021enabling,tian2021blockchain,lin2020blockchain,kumar2021privacy,zhang2020ait,zhang2020blockchain,otoum2021securing,lu2019blockchain,zhang2021bc,liu2021blockchain}. TABLE~\ref{tab:AnalysisBlockchainAIServices} presents an analysis of the integration of blockchain and AI for secure services.

\subsubsection{Spectrum Management} \label{subsubsec:SpectrumManagement}
Radio spectrum resources are scarce resources. Spectrum is widely used by various radio technologies and services, resulting in increasing demands for spectrum resources in various industries and fields. Facing ever-increasing demands for radio spectrum, spectrum management has never been more challenging. Given that traditional fixed spectrum allocation strategies lead to inefficient spectrum usage, dynamic spectrum management is proposed as an encouraging approach to alleviate the spectrum scarcity problem \cite{liang2020dynamic}. Blockchain and AI are two promising enabling technologies for solving spectrum management issues. Blockchain can be applied to spectrum auctions, which improves security and decentralization, and reduces spectrum management costs. On the other hand, AI technologies represented by DL and DRL are very powerful and can automatically learn user behavior patterns or further make optimal decisions for users. Through the use of AI, behaviors such as users' mobility and data/computing traffic can be dynamically predicted, so as to optimize the allocation of wireless resources.

In 5G beyond and 6G wireless communications, the authors of \cite{hu2021blockchain} designed a blockchain- and AI-supported dynamic resource sharing architecture. The low-cost and low-complexity hierarchical blockchain is an enabling platform for dynamic resource sharing. AI is employed to optimize data management in the process of dynamic resource sharing. The proposed architecture in \cite{hu2021blockchain} can successfully implement dynamic spectrum sharing. Simulation results show that DRL can effectively maximize the user's profit margin compared to the traditional Q-Learning algorithm and stochastic decision-making. There is a competitive relationship between multiple operators. Therefore, the operator's spectrum utilization rate and infrastructure deployment efficiency are deeply low. 6G networks will enable more flexible mobile network deployments through spectrum and infrastructure sharing among operators. Under the 6G network of multiple mobile operators, the work in \cite{maksymyuk2022ai} developed a blockchain- and AI-empowered multi-plane framework for open spectrum and infrastructure sharing. The developed framework of \cite{maksymyuk2022ai} consists of user plane, infrastructure plane, operator plane, blockchain plane, and AI plane. As a case study of the developed multi-plane framework, the authors of \cite{maksymyuk2022ai} utilized deep RNN and blockchain technology to design a workflow for dynamic spectrum management among multiple operators. Simulation results indicate that the designed intelligent dynamic spectrum management workflow can provide more equitable bandwidth allocation for all users compared with static and semi-intelligent workflows. The authors in \cite{lu2020blockchain} proposed a general edge intelligent privacy-preserving framework, which is integrated blockchain with FL and can be specifically applied to spectrum resource sharing. Spectrum sharing information is recorded in the blockchain as a transaction, and consumers pay spectrum leasing fees to providers through the blockchain. FL can not only learn computational results from data, but also provide optimized spectrum sharing strategies. In the digital twin edge network, the work of \cite{lu2020communication} provided a permissioned blockchain-based FL architecture. To improve the communication efficiency, the authors of \cite{lu2020communication} also designed an efficient asynchronous aggregation model and DRL-based algorithm to optimize user scheduling and spectrum resource allocation. For the problem of adaptive resource allocation, reference \cite{guo2019adaptive} presented a blockchain-based MEC framework, where DRL method is utilized to tackle the joint optimization problem of spectrum resource allocation and block generation. In 6G networks, reference \cite{gur2020expansive} discussed the possibility of distributed ledger technology and ML techniques to promote the coexistence of licensed and unlicensed spectrum.

\subsubsection{Computation Allocation} \label{subsubsec:ComputationAllocation}
Computing resource is one of the key resources in 6G wireless communications. Combining blockchain and AI can allocate and offload computing resources more efficiently. For instance, to facilitate the scalability and flexibility of resources, SDNIIoT, which integrates software-defined networking (SDN) into the IIoT, was proposed in \cite{qiu2018blockchain}. In large-scale distributed SDNIIoT networks, the authors of \cite{qiu2018blockchain} presented a novel permissioned blockchain-energized consensus mechanism. This mechanism synchronizes local views among different SDN controllers, and finally achieves a consensus on the global view. To further improve the throughput of the blockchain system, a joint optimization problem of view change, access selection, and computational resource allocation was constructed in \cite{qiu2018blockchain}. To this end, a dueling deep Q-learning method was proposed to deal with the joint problem. In the network environment with more and more IoT devices, reference \cite{he2020blockchain} introduced a general system framework for blockchain-assisted edge computing. In this framework, the complete procedure of transactions between IoT side and edge nodes is specified step by step. Furthermore, the authors of \cite{he2020blockchain} provided a smart contract for resource allocation in the private blockchain network. The problem of allocating edge computing resources to data service users was described as a continuous-time Markov decision process. In \cite{he2020blockchain}, the designed smart contract adopted the RL algorithm and asynchronous advantage actor-critic algorithm to tackle the problem of edge computing resource allocation. Compared with some traditional algorithms, the proposed algorithm can distinguish multiple service quality requirements of different service users, thereby ameliorating the allocation efficiency of computing resources. 

To enhance security resource management for edge users in a distributed manner, the work of \cite{manogaran2021artificial} demonstrated a blockchain-guided offloading mode, which maximizes data availability. This mode alleviates the non-probabilistic hardness problem of data availability due to cooperative and probabilistic data offloading. The data offloading process occurs in the edge network assisted by blockchain. In this data offloading mode, Naive Bayes' learning was employed to linearly classify offloaded and non-offloaded instances to obstruct service delays and unnecessary backlogs. Aiming at the security and offloading requirements in mobile edge-cloud IoT networks, the work in \cite{nguyen2021secure} developed a secure computation offloading scheme by combining blockchain and DRL. In this scheme, the computing tasks of mobile IoT devices can be offloaded to the cloud or edge servers. To upgrade the security of data offloading, a trusted smart contract-empowered access control mechanism was presented in \cite{nguyen2021secure}. This mechanism prevents cloud resources from being accessed by illegal offloading devices. Again, for example, the authors of \cite{nguyen2021secure} formulated the computation offloading, edge resource allocation, bandwidth allocation, and smart contract cost as a joint optimization problem. This joint problem can be solved using an advanced DRL algorithm with a double-duling Q-network and a optimal offloading strategy for all IoT devices can be obtained. Under the vehicular fog computing network, reference \cite{liao2020blockchain} introduced a blockchain- and ML-assisted task offloading framework. The framework utilizes smart contracts and Merkle hash trees to facilitate fair task offloading and mitigate various security attacks. Then, to tackle the task offloading optimization problem, an intelligent task offloading algorithm based on online learning was delineated in \cite{liao2020blockchain}. Without requiring the information and CSI of the vehicle fog computing server, this algorithm can learn the long-term optimal unloading strategy and effectively reduce the unloading delay, queuing delay, and switching cost. In an air-to-ground integrated power system, the work of \cite{liao2021blockchain} formulated a joint optimization problem of device-side task offloading and server-side resource allocation. Then, it demonstrated an electromagnetic interference-aware computational offloading algorithm by combining blockchain and semi-distributed learning. 

\begin{table*}[!t] 
	\centering
	\caption {Analysis of integration of blockchain and AI for secure services.} 
	\begin{tabular}{c|c|c|c|c}
		\hline \hline
		\tabincell{c}{\bf Taxonomy} &
		\tabincell{c}{\bf Representative \\ \bf References} &
		\tabincell{c}{\bf Year} &
		\tabincell{c}{\bf Key Technologies} &
		\tabincell{c}{\bf Main Contributions}
		\\ \hline 
		\multirow{12}{*}{\bf Spectrum Management} 
		&
		\multicolumn{1}{c|}{Hu et al. \cite{hu2021blockchain}} 
		&
		\multicolumn{1}{c|}{2021} 
		&
		\multicolumn{1}{c|}{\tabincell{c}{Hierarchical blockchain,\\ DRL}} 
		&
		\tabincell{l}{Proposing a blockchain- and AI-supported dynamic res-\\ ource sharing architecture in 6G and beyond networks, \\ and DRL maximizes the user's profit margin.}
		\\ \cline{2-5}             
		&
		\tabincell{c}{Maksymyuk et al. \\ \cite{maksymyuk2022ai}}
		& 
		2022
		&  
		\tabincell{c}{Blockchain, AI, \\ Deep RNN}
		&
		\tabincell{l}{A multi-plane framework based on blockchain and AI \\ for open spectrum and infrastructure sharing in the 6G \\ network with multiple mobile operators.}
		\\ \cline{2-5}    
		& 
		Lu et al. \cite{lu2020communication}
		& 
		2020
		& 
		\tabincell{c}{Permissioned blockchain, \\ Digital twin, FL, DRL}
		&
		\tabincell{l}{Presenting a permissioned blockchain-based FL archi- \\ tecture in the digital twin edge network for user sche-\\ duling and spectrum resource allocation.}
		\\ \cline{2-5} 
		& 
		Guo et al. \cite{guo2019adaptive}
		& 
		2020
		& 
		\tabincell{c}{Blockchain, MEC, \\ DRL}
		&
		\tabincell{l}{Building a blockchain-based MEC framework for ada- \\ ptive resource allocation, and DRL 
		is utilized to tackle \\ the joint optimization problem of spectrum resource \\ allocation and block generation.}
		\\ \hline 
		\multirow{14}{*}{\bf Computation Allocation} 
		&
		\multicolumn{1}{c|}{He et al. \cite{he2020blockchain}} 
		&
		\multicolumn{1}{c|}{2021} 
		&
		\multicolumn{1}{c|}{\tabincell{c}{Private blockchain, \\ RL, Edge computing}}
		&
		\tabincell{l}{Providing a general system framework for blockchain- \\ assisted edge computing and a smart contract for co- \\ mputing resource allocation in the private blockchain \\ network.}
		\\ \cline{2-5}             
		&
		\tabincell{c}{Manogaran et al. \\ \cite{manogaran2021artificial}}
		& 
		2021
		&  
		\tabincell{c}{Blockchain, Edge \\ computing, Naïve \\ Bayes' learning} 
		&
		\tabincell{l}{Demonstrating a blockchain-guided offloading mode for \\ distributed resource management of edge users, and \\ Nave Bayes' learning was employed to linearly classify \\ offloaded and non-offloaded instances.}
		\\ \cline{2-5}   
		& 
		\tabincell{c}{Nguyen et al. \\ \cite{nguyen2021secure}}
		& 
		2021
		& 
		\tabincell{c}{Blockchain, DRL, \\ Edge/cloud computing}
		&
		\tabincell{l}{Developing a secure computation offloading scheme by \\ combining blockchain and DRL for meeting the secur- \\ ity and offloading requirements in mobile edge-cloud \\ IoT networks.}
		\\ \cline{2-5}    
		& 
		Liao et al. \cite{liao2020blockchain}
		& 
		2020
		& 
		\tabincell{c}{Blockchain, ML, \\ Fog computing}
		&
		\tabincell{l}{Utilizing smart contracts and Merkle hash trees to intro- \\ duce a blockchain- and ML-assisted task offloading fra- \\ mework under the vehicular fog computing network.}
		\\ \hline 
		\multirow{12}{*}{\bf Content Caching} 
		&
		\multicolumn{1}{c|}{Qiu et al. \cite{qiu2020ai}} 
		&
		\multicolumn{1}{c|}{2020} 
		&
		\multicolumn{1}{c|}{Blockchain, DL} 
		&
		\tabincell{l}{Building a blockchain- and DL-guided edge intelligence \\ framework entitled AI-Chain, which can handle the joint \\ resource allocation problem of networking, edge compu- \\ ting and content caching.}
		\\ \cline{2-5}             
		&
		Dai et al. \cite{dai2020deep} 
		& 
		2020
		&  
		\tabincell{c}{Permissioned blockchain, \\ DRL, Edge computing}
		&
		\tabincell{l}{Combining permissioned blockchain and DRL to design \\ a secure and intelligent content caching scheme in vehi- \\ cle edge computing networks.}
		\\ \cline{2-5}   
		& 
		Cui et al. \cite{cui2020creat}
		& 
		2020
		& 
		Blockchain, FL
		&
		\tabincell{l}{Discussing a compression algorithm named CREAT ap- \\ plied to the caching, and this algorithm integrates FL \\ and blockchain.}
		\\ \cline{2-5}    
		& 
		\tabincell{c}{Zhang et al. \cite{zhang2020deep}}
		& 
		2020
		& 
		\tabincell{c}{Blockchain, DRL, \\ D2D, MEC}
		&
		\tabincell{l}{Introducing a blockchain- and smart contract-guided di- \\ stributed cache sharing incentive  mechanism to upgrade \\ user sharing-depended caching performance.}
		\\ \hline 
		\multirow{10}{*}{\bf Security and Privacy} 
		&
		\multicolumn{1}{c|}{Dhieb wt al. \cite{dhieb2020scalable}} 
		&
		\multicolumn{1}{c|}{2020} 
		&
		\multicolumn{1}{c|}{\tabincell{c}{Permissioned blockchain, \\ ML}} 
		&
		\tabincell{l}{Integrating permissioned blockchain and AI to develop \\ a distributed heterogeneous IoT network architecture \\ to add additional security	performance.}
		\\ \cline{2-5}             
		&
		Wang et al. \cite{wang2021enabling}
		& 
		2021
		&  
		\tabincell{c}{Hierarchical blockchain, \\ Transfer learning}
		&
		\tabincell{l}{A secure user authentication mechanism called ATLB \\ with the help of transfer learning and blockchain.}
		\\ \cline{2-5}   
		& 
		Kumar et al. \cite{kumar2021privacy}
		& 
		2021
		& 
		\tabincell{c}{Blockchain, DL, \\ Smart contract}
		&
		\tabincell{l}{Leveraging blockchain and DL to provide two levels of \\ security and privacy for  collaborative intelligent transp- \\ ortation systems.}
		\\ \cline{2-5}    
		& 
		Otoum et al. \cite{otoum2021securing}
		& 
		2022
		& 
		Blockchain, FL
		&
		\tabincell{l}{Constructing an adaptive trust model by combining FL\\ and blockchain, and this model treated personal trust \\ as a probability.}
		\\ \hline \hline
	\end{tabular}
	\label{tab:AnalysisBlockchainAIServices}
\end{table*}

\subsubsection{Content Caching} \label{subsubsec:ContentCaching}
Caching is introduced into the 6G communication architecture. Specifically, by deploying caching in terminals, BSs, and core network gateways, popular content is cached to the location closer to the user. Content caching can realize the local response of some user requests, reduce the transmission delay of the requested content, improve the user experience, and balance the network load. However, the current content caching strategy is relatively static. Therefore, the caching performance depends heavily on the popularity of the content and lacks the perception of users' personalized demands. At the same time, the caching deployment in the mobile environment brings great security risks to the user's data privacy. Blockchain and AI are key enabling technologies to address these challenges. For example, in \cite{qiu2020ai}, a blockchain-guided edge intelligence framework entitled AI-Chain for 6G wireless networks was pioneered, which integrates DL and blockchain. The framework benefits from the transferability of DL. Specifically, each lightweight edge node trains neural network components, and then shares local learning results on the blockchain. To demonstrate the effectiveness of the proposed framework, the work in \cite{qiu2020ai} applied AI-Chain to handle the joint resource allocation problem of networking, edge computing, and content caching. Aiming at the privacy leakage problem in cognitive vehicle networks, reference \cite{qian2020blockchain} provided a blockchain-inspired content caching architecture. Under this framework, road side units (RSUs) and vehicles that provide content cache the content in advance and broadcast it to surrounding vehicles. Vehicles with content requirements can selectively download the related content. Once the content transaction is completed, the transaction record is written to the blockchain and broadcast to all RSUs and vehicles. To enhance the cache hit ratio, the cognitive engine can sense the content demands of the underlying vehicles. Then, in \cite{qian2020blockchain}, the perception data is analyzed by ML and DL algorithms, and predictive cached results are presented to RSUs and vehicles that provide content. In vehicle edge computing networks, a secure and intelligent content caching scheme by combining permissioned blockchain and DRL was demonstrated \cite{dai2020deep}. In this scheme, vehicles accomplish content caching and BSs sustain the permissioned blockchain. Furthermore, considering vehicle mobility, the work of \cite{dai2020deep} constructed a vehicle-to-vehicle content caching optimization problem, and applied the advanced DRL algorithm to obtain the optimal caching strategies. 

In machine-to-machine communication networks, a blockchain- and edge computing-enabled network framework was proposed in \cite{li2020resource}. In this architecture, edge computing improves data caching and computing capabilities, and blockchain ensures data security and efficiency. To reduce latency, the authors of \cite{li2020resource} framed the joint optimization problem of content caching, computation offloading, and blockchain scheduling as a discrete Markov decision process. Then, a dueling optimization algorithm inspired by a dueling deep Q-network was adopted to solve this joint optimization problem. To polish up the file caching hit ratio, a compression algorithm named CREAT applied to the caching was provided in \cite{cui2020creat}. CREAT integrates FL and blockchain, which can cache files by predicting the popularity of different files through the FL algorithm and speed up the response to file requests from IoT devices. In the meantime, blockchain technology ensures the security of data transmitted by IoT devices and gradients uploaded by edge nodes. Additionally, an advanced compression algorithm is adopted in \cite{cui2020creat} to compress the uploaded gradients, so as to speed up the training process of FL. The most critical reason for the hindered development of user sharing-depended caching solutions is the lack of incentive mechanism. To upgrade user sharing-depended caching performance, the work of \cite{zhang2020deep} provided a blockchain- and smart contract-guided distributed cache sharing incentive mechanism. In this mechanism, D2D and MEC caching nodes incentivize their cache sharing willingness by receiving expected rewards. Then, to depress consensus latency and undertake confidence, a partially PBFT consensus protocol was suggested. Furthermore, both the cache placement problem and the scene selection problem can be described as Markov decision processes \cite{zhang2020deep}. Then, the DRL algorithm with deep Q-Network was presented to deal with these problems.

\subsubsection{Security and Privacy} \label{subsubsec:SecurityPrivacy}
Security and privacy vulnerabilities increase with the scale of wireless communication systems. Accordingly, the security and privacy of communication participants have become an important issue. The combination of blockchain and AI can provide more effective solutions to security and privacy challenges 6G in wireless communication systems \cite{waheed2020security}. As an illustration, a distributed heterogeneous IoT network architecture by integrating permissioned blockchain and AI was designed in \cite{dhieb2020scalable}, so as to add additional security performance. Here, blockchain is applied to share and store data of IoT devices. On the hand, ML algorithms can detect Malware and cyberattacks of distributed IoT networks and can classify these anomalous behaviors in real time. On the other hand, the existing identity authentication mechanisms have the problems of singleness and poor adaptability. Based on this consideration, the authors of \cite{wang2021enabling} introduced a secure user authentication mechanism called ATLB with the help of transfer learning and blockchain. ATLB described layered blockchain to implement the privacy protection of the authentication mechanism with collusion attack and Sybil attack. In addition, to reduce the model training time, reference \cite{wang2021enabling} added transfer learning to optimize the authentication mechanism and build a trustworthy and intelligent blockchain. In the edge service network of IIoT, a distributed ML scheme guided by blockchain can guarantee the security and privacy of data processing of multiple resource-constrained devices \cite{tian2021blockchain}. To reduce the response delay of edge services, the proposed scheme of \cite{tian2021blockchain} employed blockchain to replace cloud servers as trusted third-party institutions. Moreover, a smart contract-based incentive mechanism was applied to encourage multiple devices to participate in computing tasks. Furthermore, a size-weighted aggregation strategy was discussed to validate and integrate model parameters, thereby improving model accuracy. The SM2 public key cryptosystem was applied in \cite{tian2021blockchain} to complete the privacy protection of model parameters in edge services. There are also security and privacy challenges for data in the software-defined Internet of Vehicles. To address these challenges, the work in \cite{lin2020blockchain} developed a spatial crowdsourcing framework guided by multiple blockchains and DRL together. In \cite{kumar2021privacy}, blockchain and DL provided two levels of security and privacy for collaborative intelligent transportation systems. For the first level, smart contracts were employed for secure and intelligent data communication. For the second level, to prevent cyber attacks, LSTM auto-encoders in DL encoded data into new formats. To mitigate the precipitately growing security challenges of in-vehicle networks, a novel blockchain- and AI-empowered trust management architecture was developed in \cite{zhang2020ait}.

More interesting, many works \cite{zhang2020blockchain,otoum2021securing,lu2019blockchain,zhang2021bc,liu2021blockchain} have proved that the combination of blockchain and FL can greatly enhance data security and privacy. For example, to defeat the privacy leakage problem in traditional device failure detection, the blockchain-authorized FL scheme of \cite{zhang2020blockchain} can verify the integrity of the data. Then, an innovative centroid distance-weighted joint averaging algorithm can alleviate the data heterogeneity problem in device fault detection. In \cite{otoum2021securing}, an adaptive trust model was constructed by combining blockchain and FL, which treated personal trust as a probability. Moreover, under the constraints of certain security standards, the trust value of the terminal devices in different networks was evaluated. The study in \cite{lu2019blockchain} took advantage of blockchain and FL to design a distributed multi-party collaborative data sharing scheme. This scheme built and shared data models through FL without directly displaying the original data, thus realizing data privacy protection. Furthermore, permissioned blockchains supported secure data retrieval, thereby further controlling access to shared data and depressing the hazard of data disclosure. Similarly, integrating blockchain and FL can also be applied to enhance the security and privacy of data during data transmission \cite{zhang2021bc} and vehicle intrusion detection \cite{liu2021blockchain}.

\subsection{IoT Smart Applications} \label{subsec:6GIoTSmartApplications}
IoT has become a fundamental component of future wireless communication networks. Various smart IoT applications have great potential to provide exciting services, which is receiving more and more attention from academia and industry. IoT is a huge network formed by combining various information sensing devices with the Internet, realizing the interconnection of people, machines, and things at any time and any place. In this subsection, we will extensively discuss some important 6G IoT smart applications supported by both blockchain and AI, including smart healthcare \cite{bhattacharya2019bindaas,al2021reinforcement,veeramakali2021intelligent,kumar2021blockchain,otoum2021preventing,mallikarjuna2021blockchain,guo2021smartphone,gupta2020bats}, smart transportation\cite{al2021enabling,hassija2020traffic,tiba2020secure,song2020blockchain,zhang2019blockchain,jiang2020intelligent,fu2020autonomous,pokhrel2020federated,lu2020blockchainempowered,chai2020hierarchical}, smart grid \cite{kumar2020distributed,keshk2019privacy,wang2020aebis,ferrag2019deepcoin,jamil2021peer,gao2021fogchain,li2021integration}, and UAV\cite{singh2020deeplearning,feng2021blockchainempowered,asheralieva2019distributeddynamic,pokhrel2021blockchainbrings,islam2021blockchainbased,gumaei2021deeplearningand,aftab2020blockml,pokhrel2020federatedlearningmeets}. The analysis of integration of blockchain and AI for IoT smart applications is shown in TABLE~\ref{tab:AnalysisBlockchainAIApplications}.

\subsubsection{Smart Healthcare} \label{subsubsec:SmartHealthcare}
At present, due to the impact of the new crown pneumonia epidemic, the integration of emerging technologies and medical scenarios is accelerating. New models such as telemedicine and intelligent pre-diagnosis have become rigid needs. Thus, the smart medical industry has ushered in a new round of outbreaks. The fusion of two emerging technologies, blockchain and AI, achieves leap-forward development of smart healthcare in areas including electronic health records, health insurance, biomedical research, drug supply, procurement process management, and medical education. For instance, a multi-party electronic health record sharing framework entitled BinDaas was proposed in \cite{bhattacharya2019bindaas}. This framework integrates two technologies, blockchain and DL. Here, the blockchain stores the patient's electronic health record data in a secure manner.  DL provides future disease risk predictions for patients based on past repositories. Also, a lattice-based key and signature verification method was developed in \cite{bhattacharya2019bindaas} to fight against quantum and collusion attacks. The authors of \cite{al2021reinforcement} designed a distributed secure e-health architecture named Healthchain-RL by combining blockchain and DRL. The blockchain in the designed architecture assembled heterogeneous healthcare institutions with dissimilar demands. At the same time, the configuration of the blockchain network was optimized in real time through the online enlightened policy-making DRL algorithms, so as to accomplish a balance between security, delay, and cost. The work in \cite{veeramakali2021intelligent} investigated a blockchain- and DL-enabled secure and intelligent healthcare diagnosis scheme. This healthcare diagnosis scheme mainly involves three main steps: 1) sharing medical images based on orthogonal particle swarm optimization (OPSO) algorithm; 2) running hash value encryption through neighborhood indexing sequence algorithm; 3) performing medical diagnosis by using OPSO-DNN algorithms. 

The study of \cite{kumar2021blockchain} leveraged the advantages of blockchain and FL to constitute a detection model for computed tomography (CT) scans of COVID-19 patients. The constituted model identified COVID-19 patients from lung CT images by applying the capsule network-supported segmentation and classification method. Therein, a global DL model was trained from data collected from different hospitals and facilities using FL algorithms and blockchain was used to authenticate data from different sources. Likewise, for screening and monitoring the COVID-19 epidemic, the authors in \cite{otoum2021preventing} developed a distributed collaborative healthcare architecture guided by blockchain and FL. Different from the detection of CT images in \cite{kumar2021blockchain}, reference \cite{mallikarjuna2021blockchain} integrated blockchain and DNN to extract feature data from existing datasets, thereby helping to diagnose severe diseases such as COVID-19 and blood cancer. Detecting infectious diseases is difficult in remote and resource-poor rural areas. Meanwhile, smartphones are predicted to be one of the main tools driving improvements in healthcare delivery. Therefore, an end-to-end DeoxyriboNucleic Acid (DNA) diagnostic platform based on smartphones was proposed in \cite{guo2021smartphone}. In this platform, DL provided automatic detection of infectious disease DNA molecular test results and their analysis. Blockchain was used for secure data connection and management, thereby increasing the credibility of the entire diagnostic platform. Most critically, the authors of \cite{guo2021smartphone} also verified the feasibility of the proposed platform through field tests in rural areas. AI algorithms can predict the type of disease and surgery based on the patient's basic symptoms and historical health records. For example, the  extreme gradient boosting algorithm was applied in \cite{gupta2020bats} to classify diseases. In this reference, it described a blockchain- and AI-enabled
drone-aided smart telesurgery architecture called BATS, which introduced smart contracts to maintain the integrity and reliability of data stored on the blockchain. During emergencies in traffic jams, UAVs can transport some light healthcare items, such as medicines and surgical tools.

\begin{table*}[!t] 
	\centering
	\caption {Analysis of integration of blockchain and AI for IoT smart applications.} 
	\begin{tabular}{c|c|c|c|c}
		\hline \hline
		\tabincell{c}{\bf Taxonomy} &
		\tabincell{c}{\bf Representative \\ \bf References} &
		\tabincell{c}{\bf Year} &
		\tabincell{c}{\bf Key Technologies} &
		\tabincell{c}{\bf Main Contributions}
		\\ \hline 
		\multirow{12}{*}{\bf Smart Healthcare} 
		&
		\multicolumn{1}{c|}{\tabincell{c}{Bhattacharya et al. \\ \cite{bhattacharya2019bindaas}}} 
		&
		\multicolumn{1}{c|}{2021} 
		&
		\multicolumn{1}{c|}{\tabincell{c}{Blockchain, DL}}
		&
		\tabincell{l}{Using blockchain and DL to propose a multi-party electronic he-\\ alth record sharing framework entitled BinDaas.}
		\\ \cline{2-5}             
		&
		\tabincell{c}{\tabincell{c}{Al-Marridi et al. \\ \cite{al2021reinforcement}}}
		& 
		2021
		&  
		\tabincell{c}{Blockchain, DRL}
		&
		\tabincell{l}{Designing a distributed secure e-health architecture named Heal- \\ thchain-RL by combining blockchain and DRL.}
		\\ \cline{2-5}   
		& 
		Otoum et al. \cite{otoum2021preventing}
		& 
		2021
		& 
		\tabincell{c}{Blockchain, FL}
		&
		\tabincell{l}{A distributed collaborative healthcare architecture guided by blo-\\ ckchain and FL for screening and monitoring COVID-19.}
		\\ \cline{2-5}    
		& 
		\tabincell{c}{Mallikarjuna et al. \\ \cite{mallikarjuna2021blockchain}}
		& 
		2021
		& 
		\tabincell{c}{Blockchain, DNN}
		&
		\tabincell{l}{Integrating blockchain and DNN to extract feature data from ex- \\ isting datasets, thereby helping to diagnose severe diseases such \\ as COVID-19 and blood cancer.}
		\\ \cline{2-5}    
		& 
		Guo et al. \cite{guo2021smartphone} 
		& 
		2021
		& 
		\tabincell{c}{Blockchain, DL}
		&
		\tabincell{l}{An end-to-end DNA diagnostic platform based on smartphones \\ for driving improvements in healthcare delivery.}
		\\ \cline{2-5}    
		& 
		Gupta et al. \cite{gupta2020bats}
		& 
		2021
		& 
		\tabincell{c}{Blockchain, AI, \\ Smart contract}
		&
		\tabincell{l}{Describing a blockchain- and AI-enabled drone-aided smart tel- \\ esurgery architecture called BATS.}
		\\ \hline 
		\multirow{8}{*}{\bf Smart Transportation} 
		&
		\multicolumn{1}{c|}{\tabincell{c}{Al Ridhawi et al. \\ \cite{al2021enabling}}}
		&
		\multicolumn{1}{c|}{2021} 
		&
		\multicolumn{1}{c|}{\tabincell{c}{Blockchain, RL}} 
		&
		\tabincell{l}{A collaborative service composition	approach by combining bl-\\ ockchain and RL to improve the quality of service for vehicles.}
		\\ \cline{2-5}             
		&
		Song et al. \cite{song2020blockchain}
		& 
		2020
		&  
		\tabincell{c}{Blockchain, DNN}
		&
		\tabincell{l}{Providing a blockchain- and DNN-assisted smart vehicle co-lo- \\ calization scheme.}
		\\ \cline{2-5}   
		& 
		Jiang et al. \cite{jiang2020intelligent}
		& 
		2020
		& 
		\tabincell{c}{Blockchain, DRL, \\ Edge computing}
		&
		\tabincell{l}{Combining blockchain and multi-access edge computing to bui- \\ ld a video analytics architecture in autonomous driving systems.}
		\\ \cline{2-5}    
		& 
		\tabincell{c}{Pokhrel et al. \\ \cite{pokhrel2020federated}}
		& 
		2020
		& 
		\tabincell{c}{Blockchain, FL, \\ Consensus mechanism}
		&
		\tabincell{l}{Using blockchain-enhanced FL to propose a fully decentralized \\ communication system for autonomous vehicles.}
		\\ \hline 
		\multirow{12}{*}{\bf Smart Grid} 
		&
		\multicolumn{1}{c|}{Keshk et al. \cite{keshk2019privacy}} 
		&
		\multicolumn{1}{c|}{2020} 
		&
		\multicolumn{1}{c|}{\tabincell{c}{Blockchain, DL, \\ LSTM}} 
		&
		\tabincell{l}{Demonstrating an advanced privacy-preserving scheme by inte- \\ grating blockchain and DL in the environment of smart power.}
		\\ \cline{2-5}             
		&
		Wang et al. \cite{wang2020aebis}
		& 
		2020
		&  
		\tabincell{c}{Blockchain, FL}
		&
		\tabincell{l}{A power management system for electric vehicles merging blo- \\ ckchain and AI on the platform of smart grid.}
		\\ \cline{2-5}   
		& 
		Ferrag et al. \cite{ferrag2019deepcoin}
		& 
		2020
		& 
		\tabincell{c}{Blockchain, DL, \\ RNN}
		&
		\tabincell{l}{Designing a blockchain- and DL-guided reliable energy excha- \\ nge architecture called DeepCoin to protect the smart grid \\ from malicious attacks.}
		\\ \cline{2-5}    
		& 
		Jamil et al. \cite{jamil2021peer}
		& 
		2021
		& 
		\tabincell{c}{Blockchain, ML, \\ RNN, LSTM}
		&
		\tabincell{l}{Providing a blockchain- and ML-assisted scheme for forecast- \\ ing P2P energy transactions to effectuate real-time scheduling \\ of energy in microgrids.}
		\\ \cline{2-5}    
		& 
		Gao et al. \cite{gao2021fogchain}
		& 
		2021
		& 
		\tabincell{c}{Blockchain, Edge-AI}
		&
		\tabincell{l}{Adopting blockchain and edge-AI to formulate distributed ener- \\ gy trading and management
			system for smart microgrids named \\ FogChain.}
		\\ \hline 
		\multirow{8}{*}{\bf UAV} 
		&
		\multicolumn{1}{c|}{Singh et al. \cite{singh2020deeplearning}} 
		&
		\multicolumn{1}{c|}{2021} 
		&
		\multicolumn{1}{c|}{\tabincell{c}{Blockchain, DL}} 
		&
		\tabincell{l}{Introducing a security scheme for information transmission be- \\ tween UAVs, which integrated blockchain and DL.}
		\\ \cline{2-5}             
		&
		Feng et al. \cite{feng2021blockchainempowered}
		& 
		2022
		&  
		\tabincell{c}{Blockchain, FL}
		&
		\tabincell{l}{A novel secure identity authentication approach by leveraging \\ blockchain-backed FL to overcome the security challenges \\ of cross-domain UAVs' authentication.}
		\\ \cline{2-5}   
		& 
		\tabincell{c}{Pokhrel et al. \\ \cite{pokhrel2021blockchainbrings}}
		& 
		2021
		& 
		\tabincell{c}{Blockchain, FL}
		&
		\tabincell{l}{A blockchain- and FL-assisted knowledge sharing and collabo- \\ rative learning scheme for UAV swarms or LEO satellites.}
		\\ \cline{2-5}    
		& 
		\tabincell{c}{Gumaei et al. \\ \cite{gumaei2021deeplearningand}}
		& 
		2021
		& 
		\tabincell{c}{Blockchain, DNN, \\ Edge computing}
		&
		\tabincell{l}{Presenting a UAV recognition and detection architecture by co- \\ mbining blockchain, deep DNN, and edge computing.}
		\\ \hline \hline
	\end{tabular}
	\label{tab:AnalysisBlockchainAIApplications}
\end{table*}

\subsubsection{Smart Transportation} \label{subsubsec:SmartTransportation}
With the rapid development of information technology, smart transportation has also ushered in more development opportunities. Smart transportation refers to the full use of big data, IoT, cloud computing, blockchain, AI, and other technologies in the field of transportation. Smart transportation can fully guarantee traffic safety, give full play to the efficiency of transportation infrastructure, and improve the operational efficiency and management level of the transportation system. In this part, we focus on the role of blockchain and AI in promoting smart transportation. To improve the quality of service for vehicles in 6G networks, the study in \cite{al2021enabling} discussed a collaborative service composition approach by combining blockchain and RL. Here, the blockchain was applied to announce combined tasks under certain constraints of service requests, ensuring that adjacent nodes interact securely and record transactions. To quicken the procedure of service composition path choice, the authors of \cite{al2021enabling} employed an RL algorithm to pick the optimal solution closest to the node request. To ward off traffic congestion, reference \cite{hassija2020traffic} described a blockchain-guided secure crowdsourcing scheme. This scheme encouraged users to voluntarily take part in traffic forecasting by sharing traffic information to earn tokens. Meanwhile, users can also spend these tokens to acquire the required traffic information from the network. Then, with the help of an LSTM neural network, the study of \cite{hassija2020traffic} fused the results of a feed-forward artificial neural network trained on historical data to prognosticate traffic congestion probabilities on real-time data. Similarly, the work in \cite{tiba2020secure} considered integrating blockchain, RL, and edge computing to alleviate the traffic congestion problem.

Currently, vehicle positioning has challenges of low accuracy and network congestion in data sharing. To address these challenges, a blockchain- and DNN-assisted smart vehicle co-localization scheme was provided in \cite{song2020blockchain}. This scheme was benchmarked with multiple traffic signs, and the DNN algorithm was applied to correct position of vehicles. For the localization errors of multiple vehicles, the work of \cite{song2020blockchain} demonstrated a DL-inspired method for distance computation and prognostication to turn down localization errors for common vehicles on the unchanged road segment. Additionally, to actualize the safety information sharing between vehicles, the corresponding mechanisms of message asking, message selecting, message sharing, and punishment mechanism were presented based on smart contracts. In the in-vehicle self-organizing environment, an advanced blockchain-authorized distributed software-defined security architecture was delineated \cite{zhang2019blockchain}. Then, the dueling deep Q-learning algorithm with prior experience playback was adopted in \cite{zhang2019blockchain} to procure the optimal strategy while satisfying the requirement of maximizing the system throughput. In autonomous driving vehicle networks, the sharing and storage of massive video data is terribly difficult. To cope with these difficulties, reference \cite{jiang2020intelligent} combined blockchain and multi-access edge computing to build a video analytics architecture in autonomous driving systems. This architecture completed the secure storage and sharing of video data with the help of smart contract. Then, the study of \cite{jiang2020intelligent} formulated the joint optimization problem of video offloading and resource allocation as a Markov decision process. Moreover, a high-level DRL algorithm with asynchronous advantage actor-critic was proposed to tackle this joint problem. Improper lanes for self-driving cars can cause tragic accidents, and thus, the authors of \cite{fu2020autonomous} demonstrated an autonomous lane changing system assisted by blockchain and collective learning. Blockchain ensured data security for autonomous vehicles while encouraging vehicle resources to join in collective learning. Then, an advanced algorithm based on the deep deterministic policy gradient was used in \cite{fu2020autonomous} to address the lane changing problem, so as to achieve optimal autonomous driving policies.

Interestingly, the fusion of blockchain and FL powerfully polishes up the performance of intelligent transportation systems \cite{pokhrel2020federated,lu2020blockchainempowered,chai2020hierarchical}. For instance, on the basis of blockchain-enhanced FL, the study in \cite{pokhrel2020federated} proposed a fully decentralized communication system for autonomous vehicles. In the proposed system, the local on-vehicle ML model updates are interchanged and validated with other vehicles in the distributed manner. At the same time, the proposed system made full use of the consensus mechanism of the blockchain and can complete the update of local vehicle ML models without any third-party server. To enhance the reliability of edge data sharing between vehicles, a hybrid blockchain- and FL-assisted secure data sharing framework was designed in \cite{lu2020blockchainempowered}. In this framework, the hybrid blockchain was composed of the permissioned blockchain and the local directed acyclic graph of vehicle operation, which can further upgrade the security of in-vehicle data. Furthermore, to further polish up the training efficiency, the authors of \cite{lu2020blockchainempowered} introduced an asynchronous FL scheme for model learning from edge data, and preferred the better participating node through the DRL algorithm. The work of \cite{chai2020hierarchical} demonstrated a layered blockchain- and FL-embedded vehicle knowledge sharing system. This system adopted a hierarchical blockchain to record the FL model. Then, integrating a proof-of-learning-based consensus protocol with high-precision hierarchical FL effectively prevent the waste of a large amount of computing resources. Also, the knowledge sharing process among vehicles was constructed as a multi-leader and multi-follower non-cooperative game problem in \cite{chai2020hierarchical}.

\subsubsection{Smart Grid} \label{subsubsec:SmartGrid}
The construction of a smart grid provides a strong guarantee for improving the related functions of smart cities, thus further accelerating the pace of urban intelligence. As a network with extensive coverage, the smart grid should realize the interaction with users, the intelligentization of power grid equipment, the full automation of power production, and the greening of energy, so as to comprehensively improve the level of informatization and intelligence of the power grid. Modern communication technology is fully utilized to build a safe, reliable, green, and efficient smart grid. The application of blockchain and AI to the smart grid can improve the quality and efficiency of grid engineering, thereby enhancing the stability of the smart grid system. For example, the work of \cite{kumar2020distributed} systematically discussed the enabling role of blockchain, AI and IoT in improving smart grid performance. In the environment of smart power, reference \cite{keshk2019privacy} demonstrated an advanced privacy-preserving scheme by integrating two emerging technologies, blockchain and DL, which consisted of a two-level privacy mechanism and an anomaly detection mechanism. Specifically, the first-level privacy mechanism applied a blockchain found on an enhanced PoW consensus protocol to validate data integrity and diminish data poisoning attacks. To avoid inference attacks, the second-level privacy mechanism converted the raw data into an encoded form with the assistance of a variational autoencoder. Then, in \cite{keshk2019privacy}, an LSTM-inspired anomaly detection mechanism employed two public datasets to train and validate the output of the two-level privacy mechanism. On the platform of smart grid, a power management system for electric vehicles merging blockchain and AI was creatively demonstrated in \cite{wang2020aebis}. The system applied artificial neural networks and FL to prophesy the electricity consumption of electric vehicles. At the same time, the blockchain can incorporate all distributed electric vehicles to form a smart energy storage framework. The blockchain traded memory and time for the security and transparency performance of the proposed power management system. To protect the smart grid from malicious attacks, the study of \cite{ferrag2019deepcoin} designed a blockchain- and DL-guided reliable energy exchange architecture called DeepCoin. In DeepCoin, the blockchain adopted the PBFT algorithm to accomplish the consensus in the P2P energy system, thereby promoting users to voluntarily trade redundant energy to other adjacent users. In \cite{ferrag2019deepcoin}, the blockchain also applied bilinear pairing, short signature, and hash function to complete the privacy protection of smart grid users. Then, DeepCoin tracked down cyber-attacks and deceitful transactions in smart grids through the RNN-based DL algorithm. 

Recently, microgrids are small-scale power systems that utilize renewable energy sources to distribute electricity near users. To effectuate the real-time scheduling of energy in microgrids, the authors of \cite{jamil2021peer} provided a blockchain- and ML-assisted scheme for forecasting P2P energy transactions. This energy transaction scheme was modeled and completed on the permissioned blockchain network entitled Hyperledger Fabric. At the same time, smart contracts performed real-time scheduling of distributed energy and controllable loads. In addition, RNN, LSTM, and bidirectional LSTM-powered ML algorithms are employed to forecast energy requirements in microgrids while downgrading electricity transportation expenses. In \cite{gao2021fogchain}, a distributed energy trading and management system for smart microgrids named FogChain was presented. FogChain adopted blockchain to formulate a decentralized energy trading platform and applied edge-based AI methods to draw distributed controllers for microgrids. In the same direction, the work in \cite{li2021integration} combined blockchain and ML to address data sharing, processing, and forecasting problems in microgrid systems.

\subsubsection{UAV} \label{subsubsec:UnmannedAerialVehicles}
In recent years, with the development of information technology, UAV has formed an intelligent aircraft that combines multiple technologies such as flight control, network communication, and electric power. UAVs can be regarded as flying IoT devices, and have been widely applied in military, agriculture, forestry, transportation, meteorology, and other fields. UAV has the advantages of low cost, high dynamics, and deployment flexibility. However, UAV communication faces many threats such as being susceptible to interference, inability to cover large areas, and unstable communication. The combination of blockchain and AI provides new research ideas for alleviating these threats. For example, under the UAV Internet system, reference \cite{singh2020deeplearning} introduced a security scheme for information transmission between UAVs, which integrated blockchain and DL. This scheme applied a zero-knowledge-proof-based blockchain to maintain the security and privacy of data dissemination between UAVs. Moreover, the DL-inspired miner selection algorithm in \cite{singh2020deeplearning} can obtain the optimal miner node strategy, thereby shortening the block generation time and transaction submission time. To overcome the security challenges of cross-domain UAVs' authentication, a novel secure identity authentication approach by leveraging blockchain-backed FL was provided in \cite{feng2021blockchainempowered}. In this approach, FL only shared the data model uploaded by the authenticated UAV  instead of directly sharing the original data. The authors of \cite{feng2021blockchainempowered} made full use of multi-signature smart contracts to practice distributed cross-domain UAVs' identity authentication. And, to surmount the single point failure, these multi-signature smart contracts were also employed to perform aggregations of global model updates. In the IoT environment, the work of \cite{asheralieva2019distributeddynamic} proposed a distributed dynamic resource management and pricing system that integrated blockchain-as-a-service (BaaS) and UAV-authorized MEC. Specifically, MEC servers are installed on both the ground BSs and the UAVs acting as the air BSs to process some blockchain tasks. BaaS combined blockchain and cloud computing so that resource management and pricing can be handled on BSs with MEC servers in \cite{asheralieva2019distributeddynamic}. Then, in the case of incomplete information, the interaction process of resource management and pricing between BSs and peers of the proposed system was expressed as a stochastic Stackelberg game with multiple leaders.

Additionally, UAV swarms or low earth orbit (LEO) satellites are extremely vulnerable to security threats. Therefore, reference \cite{pokhrel2021blockchainbrings} elucidated a blockchain- and FL-assisted knowledge sharing and collaborative learning scheme. Specifically, this scheme considered the influence of the number of miners, block transfer, and the mobility of UAVs/LEOs, the authors of \cite{pokhrel2021blockchainbrings} derived the probability of regular forks and optimized the energy consumption of PoW computation for blocks. More importantly, in \cite{pokhrel2021blockchainbrings}, an advanced FL-enabled algorithm was used to complete the resource allocation of mobile mining. And, the coordination gains of blockchain and FL for UAV swarms was illustrated. The work of \cite{islam2021blockchainbased} integrated blockchain, AI, and UAV swarms to design an autonomous detection framework for infectious diseases. Here, UAV swarms can expand coverage and  lessen human participation. This detection framework applied a lightweight blockchain and two-stage security authentication mechanism to remote areas where the network is scarce to degrade the burden of UAVs. Then, a DL-inspired algorithm was provided to autonomously detect disease prevalence in \cite{islam2021blockchainbased}. The security of RF signal transmission between UAVs and the accuracy of identification and detection are challenged. Consequently, a UAV recognition and detection architecture by combining blockchain, deep DNN, and edge computing was elucidated in \cite{gumaei2021deeplearningand}. This architecture applied blockchain to protect the security of data transmission. Moreover, the deep DNN was adopted for training by using the collected RF signal data from UAVs in different flight modes. Then, the trained model was downloaded to edge devices to identify UAVs and detect their flight modes. There are trade-off problems in terms of quantity, energy consumption, coverage area, and height when installing BSs on the UAV side (UAV-BS). To address this deployment problem of UAV-BS, the study of \cite{aftab2020blockml} demonstrated a blockchain- and ML-guided smart placement scheme for UAV-BS. More interestingly, in 6G networks, the work in \cite{pokhrel2020federatedlearningmeets} combined blockchain and FL can provide a new idea for UAV-assisted construction of disaster response systems.

\subsection{{Analysis of Operating Frequencies, Visions, and Requirements from the 6G Perspective}} \label{subsec:Analysis6GPerspective}
\subsubsection{{Operating Frequencies from the 6G Perspective}}\label{subsubsec:OperatingFrequencies}
{{ Considering the operating frequencies of AI and blockchain for 6G is crucial from 6G perspective, as 6G networks have higher requirements for high-speed data transmission and processing. However, there are currently no established standards or fixed ranges for the operating frequencies of AI and blockchain in 6G. Since 6G technology is still in the research and standardization phase, there is limited discussion and research on this specific topic in the existing literature. Current works \cite{yang20196g, zhang20196g,nguyen20216g,saad2019vision,chowdhury20206g,wang2023road,de2021survey} primarily focus on the communication characteristics, spectrum range, and key technologies of 6G. The operating frequencies in 6G are expected to encompass a wide range of spectrum, including low-frequency, mid-frequency, high-frequency, as well as millimeter-wave and terahertz bands. However, the current work does not specifically address the operating frequencies of AI and blockchain in 6G. Until further research and standardization developments, specific discussions and research on the operating frequencies of AI and blockchain for 6G may remain limited. 

Although specific references may be limited, we can provide a general discussion on the operating frequencies of AI and blockchain in 6G networks. \textbf{Millimeter-wave and Terahertz Bands:} 6G networks will utilize high-frequency millimeter-wave and terahertz bands to achieve higher data transmission rates and lower latency. The use of these frequency bands will impact the operating frequencies of AI and blockchain, requiring consideration of their performance and adaptability in these bands. \textbf{Power Consumption and Resource Management:} 6G networks demand high performance while minimizing power consumption. When determining the operating frequencies of AI and blockchain, a comprehensive consideration of power consumption and resource management is necessary to achieve efficient computation and communication. \textbf{Adaptive Adjustment Strategies:} Given the dynamic nature and diverse application requirements of 6G networks, adaptive adjustment strategies are crucial for the operating frequencies of AI and blockchain. By continuously monitoring and analyzing network conditions, application demands, and resource availability, it is possible to dynamically adjust the operating frequencies of AI and blockchain to meet real-time requirements and optimize network performance.

The operating frequencies of AI and blockchain for 6G will be determined based on specific application scenarios and requirements. For example, in the fields of IoT, UAV or smart cities, different frequency bands may be used to support the 6G secure services and 6G IoT smart applications by integrating of AI and blockchain. The specific operating frequencies will depend on communication requirements, device characteristics, as well as the needed bandwidth and capacity. Therefore, from the 6G perspective, determining the operating frequencies of AI and blockchain requires considering multiple factors and referencing the development of future 6G standards and the demand of practical application. Notably, the discussion provided above only offers general perspectives on operating frequencies and does not establish specific frequency band ranges. As 6G technology continues to be researched and developed, future studies and standardization efforts will provide more specific and detailed guidance.
}}

\subsubsection{{Visions from the 6G Perspective}}\label{subsubsec:Vision}
{{The vision of 6G is to build a highly intelligent, highly connected, and highly adaptive network to meet the needs of future society and industries. From the 6G perspective, the visions include several key aspects. \textbf{Ultra-high rates and ultra-low latency:} The vision of 6G is to achieve higher data transmission rates and lower communication latency to support a wide range of applications, including enhanced mobile broadband, virtual and augmented reality, high-definition video, etc. Through high-speed and low-latency network transmission, the data processing and interaction capabilities of AI and blockchain will be enhanced, supporting more complex and real-time application scenarios. This will provide a stronger foundation for integrated AI and blockchain applications. \textbf{Super connectivity:} The vision of 6G is to establish a super-connected network that enables highly interconnected devices. The super-connected network includes D2D, device-to-infrastructure, and device-to-cloud connections. This will provide more connectivity options and broader coverage for AI and blockchain services and applications. \textbf{Powerful intelligence and adaptability:} The vision of 6G is to build an intelligent network with edge computing and distributed intelligence capabilities. This will enable real-time processing and decision-making of AI and blockchain technologies at the network edge, reducing latency and improving performance. Moreover, the intelligent network will be able to adaptively optimize based on application requirements, providing more efficient support for AI and blockchain services and applications. \textbf{Security and privacy protection:} The vision of 6G is to ensure network security and privacy protection to address the growing security threats. In the services and applications of AI and blockchain, security and privacy protection are crucial. 6G will provide stronger security mechanisms, including encryption, identity authentication, and access control, to ensure the secure and reliable operation of AI and blockchain services and applications.

In summary, the vision of 6G is an evolving and developing concept, and there is no unified definition or standard yet. Therefore, our discussion is primarily based on current research and academic discussions \cite{yang20196g, zhang20196g, saad2019vision,wang2023road} provide a general 6G vision for blockchain and AI services and applications.				
}}

\subsubsection{{Requirements from the 6G Perspective}}\label{subsubsec:Requirements}
{{To achieve the 6G vision discussed above, it naturally leads to the demand for higher bandwidth, lower latency, super connectivity, high security, and high reliability in 6G. Specifically, in the topic of blockchain and AI for 6G, the requirements of 6G include the following key aspects. \textbf{High bandwidth demand:} With the increasing adoption of AI and blockchain services and applications, there is a growing need for higher bandwidth to support large-scale data transmission, real-time decision-making, and complex computations. \textbf{Low latency requirement:} AI and blockchain services and applications require real-time responsiveness. To support these services and applications, 6G needs to provide lower communication latency to ensure fast data transmission and processing capabilities. \textbf{Large-scale connectivity capability:} With the proliferation of IoT devices and AI applications, 6G needs to have the ability to connect a massive number of devices to facilitate interconnection and data exchange. \textbf{High security and privacy protection:} Security and privacy are crucial in AI and blockchain services and applications. 6G needs to provide robust security mechanisms, including identity authentication, encryption, and secure transmission, to ensure the confidentiality and integrity of data. \textbf{High reliability and robustness:} 6G should exhibit high reliability and robustness to handle various network environments and cope with interferences and failures. This will ensure the stability and reliability of AI and blockchain services and applications.
		
In summary, the aforementioned requirements \cite{zhang20196g,chowdhury20206g,wang2023road,de2021survey} are based on the 6G perspective of integrating AI and blockchain services and applications. As 6G technology continues to evolve and standardization efforts progress, these requirements may be further refined and supplemented.

In this section, we have broadly discussed the services and applications of merging blockchain and AI for 6G networks. This includes the integration of blockchain and AI for 6G secure services, such as spectrum management, content caching, computation allocation, and security and privacy. We have also examined the 6G IoT smart applications of merging blockchain and AI, covering areas such as smart healthcare, smart transportation, smart grid, and UAV. Furthermore, we have thoroughly discussed the operating frequencies, visions, and requirements from the 6G perspective. In summary, this section highlights the complementary nature of blockchain and AI in addressing existing challenges and meeting the evolving requirements of the 6G era. The integration of blockchain and AI in 6G networks demonstrates the potential to enhance secure services and enable advanced IoT applications. Additionally, the analysis of operating frequencies, visions, and requirements provides valuable insights for shaping the future of 6G networks. By studying the integration of blockchain and AI for 6G networks, we have gained a deeper understanding of their synergistic capabilities and the opportunities they present for transforming wireless communications.  		
}}

\section{Open Issues, Research Challenges, and Future Work} \label{sec:ResearchChallengs}
Integrating blockchain and AI in 6G wireless communications is currently a hot research topic. In the previous sections, we focused on investigating the possibilities of combining blockchain and AI. Moreover, we extensively discussed the integration of blockchain and AI for wireless communications, involving secure services and IoT smart applications. Research on integrating blockchain and AI for wireless communications is still emerging, but future works need to address some questions and challenges. In this section, based on extensive research works in current literatures, we summarize the possible open issues and research challenges for integrating blockchain and AI in 6G wireless communications. The purpose is to provide beneficial inspirations and references for future innovation research. We also explore potential research directions in the future.

\subsection{Towards Blockchain} \label{subsec:TowardsBlockchain} 
In recent years, the research and application of blockchain has begun to grow explosively. Blockchain is considered to be the key technology leading the current information internet to the value internet. Although blockchain has great potential for 6G networks, there are still some challenges that limit its widespread application in 6G networks.

As the number of transactions increases significantly, scalability is the biggest hurdle limiting the widespread adoption of blockchain technology in 6G networks. For example, the Bitcoin and Ethereum systems process an average of 7 to 20 transactions per second, far behind the Visa system that handles tens of thousands of transactions per second. Moreover, the multi-copy feature of the blockchain requires a large amount of additional storage space, which increases storage costs. This will result in limited space utilization, making it difficult to support large-scale applications. There have been many researchers improving the scalability of blockchains by using techniques such as sidechains, lightning networks, sharding, pruning, and directed acyclic graphs. However, these methods still have their own problems, such as how to properly divide the tiles, and which transactions to prune. In addition, the blockchain is a high-energy-consumption industry. As an illustration, the blockchain system based on the PoW consensus mechanism relies on the computing power contributed by the blockchain nodes. However, only part of the computing power has been rewarded, and other computing power is doing useless work, which wastes a lot of resources. This problem of high energy consumption problem affects the popularization and application of blockchain in 6G. 

Security attack is the most important problem faced by the blockchain so far, such as Bitcoin's 51\% attack and botnet attack. The asymmetric encryption mechanism of the blockchain will become more and more fragile with the development of mathematics, cryptography, and computing technology. Security issues are also a great threat to the further application of blockchain to 6G networks. The data transactions recorded on the blockchain are open and transparent, which is beneficial for data sharing and verification, but not conducive to the privacy protection of user information. As more and more personal data is stored in blockchain-powered 6G networks, privacy leakage becomes another key issue.

\subsection{Towards AI} \label{subsec:TowardsAI} 
In recent years, AI, especially DL, has achieved great success in computer vision, natural language processing, speech recognition, and other fields. The researchers expect to apply AI to all levels of the 6G system, thereby generating an intelligent communication system, realizing the true interconnection of everything, and meeting people's ever-changing demands for data transmission rates. However, there are still many challenges and unsolved issues in implementing and managing complex intelligent communication systems.

Inefficient data management schemes and high overhead of information exchange among communication participants are key bottlenecks in the development of AI technology. The AI-based solutions, such as ML methods, usually require large amounts of training data, which need to be collected and implemented on a centralized server with sufficient storage and computing resources. Nevertheless, current wireless communication systems do not have access to massive amounts of data to train models. In heterogeneous networks, aggregating data from different sources to train models is also an open problem. In 6G networks, users may have different service quality requirements in different scenarios. For example, in video streaming applications, users demand high throughput and low latency at the cost of security. However, in payment softwares, users require high security, even at the expense of throughput. In this direction, designing a cross-layer, action-based AI protocol for different applications is a key issue to satisfy various service demands while balancing the network resources of 6G networks. 

Notably, in the current era of rapid development of AI, data security and privacy issues are receiving increasing attention. 6G network realizes interconnected intelligence by supporting AI functions, and adopts a centralized network architecture, which is vulnerable to hacker attacks. Moreover, 6G network needs to collect a large amount of user data for training through billions of devices. The training data involves a large amount of personal information, so AI can easily lead to privacy leakage of user data. Using distributed technology to design AI-enabled 6G network architectures can actualize a decentralized security and trustworthy mechanism. Without sending all data to the cloud computing center, the distributed technology processes data where it is generated, which can alleviate the problem of privacy leakage to a certain extent. Nevertheless, communication needs to exchange the knowledge information perceived by both parties and update the knowledge bases of both parties collaboratively, which also leads to the risk of privacy leakage of local data. How to develop an efficient coordination mechanism among communication participants without causing any private data leakage remains an urgent problem to be solved.

\subsection{Towards Blockchain- and AI-assisted Wireless Communications }
Integrating blockchain and AI brings new opportunities to 6G networks, as well as some open issues and research challenges. For example, the natural conflict between blockchain and AI, the processing of a large amounts of data, and the collaborative optimization of multiple systems and multiple indicators. Furthermore, the effectiveness and feasibility of blockchain- and AI-assisted wireless communications still need to be verified by large-scale practice of wireless communication networks.

There are some conflicts with the combination of blockchain and AI. For example, the execution results of smart contracts in blockchains are often deterministic. While, the execution results of AI algorithms are usually uncertain, random, and unpredictable in most cases. The contradiction between blockchain and AI poses certain challenges for AI embedded in blockchain to optimize the execution decisions of 6G networks. Therefore, in future research, a new solution is required to deal with the contradiction between the certainty of smart contracts and the randomness of AI algorithms. The new solution can handle approximate calculations for smart contracts and design consensus protocols for each participating node of blockchain. The purpose of the new solution is to output the decision results under the 6G network with specific certainty, high accuracy, and high precision. With the explosive growth of data in 6G networks, the processing of large amounts of data in blockchain- and AI-assisted wireless communication systems is a terribly large challenge. Typically, blockchain is applied to securely collect and store large amounts of data, and AI uses these data for model training processing. The massive unlabeled and unclassified datasets are intractable for AI training. At the same time, the blockchain also presents a potential bottlenecks in storing these large-scale distributed data. For example, the data recorded on the blockchain is open and transparent. The blockchain-based storage method is beneficial for data sharing and verification, but not conducive to data privacy protection. Furthermore, at present, most works in related kinds of literature often only optimize a single performance index in a single wireless communication system. The collaborative optimization of multiple performance indexes of multiple wireless communication systems is ignored. However, with the maturity and in-depth research of blockchain and AI technologies in the future, the solutions proposed for the performance of different wireless communication systems can be combined with each other. Thereby, the goal of co-optimizing multiple performances of multiple systems can be achieved.

Observing the existing literature, blockchain- and AI-assisted wireless communications are still in the infant stage. Many works apply blockchain to create a trusted environment for wireless communications, and provide prediction, optimization, identification, detection, and decision-making for wireless communication systems through AI algorithms. Few works have really deeply integrated blockchain and AI into wireless communications. Whether we look at the current technical indicators of blockchains or the actual implementation of AI and 6G networks, there are still many uncertainties to truly realize the integration and implementation of blockchain and AI technologies for 6G networks. The potential outcomes of the fusion of blockchain and AI for 6G networks are also difficult to assess. Therefore, while actively investigating the integration of blockchain and AI for 6G networks, we must also look at it rationally and focus on practical implementation. In the future, we will continue to take an organic combination and flexible and innovative approach to truly realize the practice and exploration of the integration of blockchain and AI in 6G wireless communications.

\subsection{Future Work}
In future work, we still need to tackle the technical barriers of blockchain and AI. We can deeply investigate the matching and joint optimization of blockchain- and AI-powered 6G networks in terms of performance indicators, security, stability, etc. We can also further research and ensure the healthy and sustainable development of blockchain and AI technologies in 6G networks. In addition, the Federal Communications Commission, at the 2018 US Mobile World Congress, emphasized that 6G can introduce blockchain technology into spectrum sharing \cite{FCCBlockchain}. Research institutions such as the Institute of Electrical and Electronics Engineers and the France's Spectrum Regulator have also begun to explore the application of blockchain to manage spectrum \cite{weiss2019application}. Thus, we will focus on the fusion of blockchain and AI to obtain smarter and more distributed dynamic wireless resource allocation. 6G can combine emerging advanced technologies such as cloud computing, edge computing, and big data to promote the development of blockchain and AI technologies. Accordingly, the mutual promotion and integration of blockchain, AI, and 6G is also one of the key research directions in the future. With the further improvement of blockchain and AI, academia and industry will continue to transform theory into technology and put it into practice. We believe that future integration of blockchain and AI will be more in-depth, and the scenarios applied to 6G networks will be more abundant. Deploying Blockchain and AI in 6G networks will bring more surprises and possibilities to our lives.

{{In this section, we have outlined the open issues and research challenges associated with integrating blockchain and AI in 6G wireless communications. Future work should focus on innovative solutions for scalability and energy efficiency in blockchain, as well as exploring techniques for security and privacy protection. Additionally, efficient data management schemes and cross-layer AI protocols need to be developed to address AI-related challenges. The integration of blockchain and AI for 6G networks is still in its early stages, and future research should aim to overcome technical barriers and uncertainties, promote mutual promotion and integration, and explore practical applications. Overall, addressing these challenges and furthering the integration of blockchain and AI in 6G networks will bring significant advancements to wireless communications.
}}

\section{Conclusion} \label{sec:ConclusionFutureWork}
This survey reviewed the latest progress of blockchain and AI for 6G wireless communications. We began our comprehensive survey with a basic overview of blockchain and AI. Specifically, we briefly described the concepts, characteristics, and categories of blockchain and AI. The recent developments in applying blockchain and AI to wireless communications, respectively, were also showcased. To thoroughly explore the possibility of combining blockchain and AI, we started with two aspects of blockchain-assisted AI and AI-aided blockchain. We also highlighted the motivations for integrating blockchain with AI for 6G wireless communications. Next, we then extensively discussed the simultaneous deployment of blockchain and AI in 6G wireless communication systems, involving secure services and IoT smart applications. In particular, a comprehensive exploration of the widely popular secure services supported by blockchain and AI was conducted, spotlighting spectrum management, computation allocation, content caching, and security and privacy services. In addition, we also covered the latest achievements of blockchain and AI empowerment in various IoT smart applications. We made an exhaustive analysis from four scenarios: smart healthcare, smart transportation, smart grid, and UAV.  {{Furthermore, we have thoroughly discussed operating frequencies, visions, and requirements from the 6G perspective.}} Finally, we have pointed out several open issues, research challenges, and potential research directions toward blockchain and AI for 6G networks.

{In summary, this survey attempts to comprehensively explore the technologies related to blockchain and AI for wireless communications. Although the research on integrating blockchain and AI for 6G networks is still in its infancy, it is obvious that blockchain and AI will significantly uplift the performance of various services and applications in 6G networks. We believe our study will shed valuable insights into the research challenges associated with blockchain- and AI-assisted 6G networks as well as motivate interested researchers and practitioners to devote more research efforts to this promising area.}

\bibliographystyle{IEEEtran}
\bibliography{IEEEabrv,Introduction,BC,AI,BC_AI,BC_AI_6G}
\end{document}